%% file: DARS-SIM-II-IEEE.tex
\documentclass[10pt,journal]{IEEEtran}
\usepackage{graphicx}
\usepackage{amsmath,amsfonts,amssymb,epsfig,epstopdf,url,array}

\usepackage{amsthm}
\usepackage{color,soul}
\usepackage{algorithmic}
\usepackage{balance}
\usepackage[ruled]{algorithm2e}
\usepackage{stmaryrd}
\usepackage[table]{xcolor}
\usepackage{centernot}
\usepackage{tikz}
\usepackage{booktabs}
\usepackage{pdfpages}
\usepackage{graphicx}
\usepackage{subfigure}
\usepackage[font=small,skip=8pt]{caption}
\usepackage{tabularx}
\usepackage{multirow}
\usepackage{multicol}
\usepackage{bigstrut}
\usepackage{pdflscape}
\usepackage{rotating}
\usepackage{float}
\usepackage{lipsum}
\usepackage{enumitem}
\usetikzlibrary{plotmarks,shapes,arrows,chains,hobby,backgrounds,calc,trees}
\usepackage{makecell}
\usepackage{dblfloatfix}    
\usepackage{arydshln}
\usepackage{amssymb}
\usepackage{bm}
\usepackage{arydshln}
\usepackage{pgfplots}
\usepackage{pgfplotstable}
\pgfplotsset{compat=newest}
\usepackage{lipsum}                     
\usepackage{xargs}                      
\usepackage[colorinlistoftodos,prependcaption,textsize=tiny]{todonotes}

\definecolor{lightgray}{gray}{0.9}
\theoremstyle{definition}
\newtheorem{exmp}{Example}
\theoremstyle{definition}
\newtheorem{mydef}{Definition}
\theoremstyle{definition}

\newcommand*{\Perm}[2]{{}^{#1}\!P_{#2}}%

\definecolor{header}{rgb}{0.0,0.0,0.0}
\definecolor{myblue}{rgb}{0.5,0.5,0.5}
\begin{document}
\title{A Fuzzy-Based Optimization Method for \\ Integrating Value Dependencies into \\ Software Requirement Selection}
\author{Davoud~Mougouei and~
        David~Powers
\IEEEcompsocitemizethanks{
	\IEEEcompsocthanksitem D. Mougouei was with the School of Computing and IT, University of Wollongong, Australia (dmougouei@gmail.com)
\IEEEcompsocthanksitem D. Powers was with Flinders University, Australia (david.powers@flinders.edu.au)
}
}
\markboth{}%
{Mougouei \MakeLowercase{\textit{et al.}}: Dependency-Aware Requirement Selection in Software Release Planning}
\IEEEtitleabstractindextext{%
\begin{abstract}
\input{abstract}
\end{abstract}
}
\maketitle
\IEEEdisplaynontitleabstractindextext
\IEEEpeerreviewmaketitle
\input{introduction}
\input{related}
\input{identificationANDmodeling}
\input{identification}
\input{modeling}

\input{selection}
\input{selection_ov}

\input{selection_ilp}

\input{selection_blind}
\input{simulation}
\input{simulation_I}

\input{simulation_II}

\input{simulation_III}

\input{simulation_IV}

\input{simulation_V}

\input{simulation_VI}
\input{scalability}
\input{conclusion}

\ifCLASSOPTIONcaptionsoff
  \newpage
\fi
\bibliographystyle{IEEEtran}
\bibliography{ref}
\end{document}

%% file: abstract.tex
Software requirement selection aims to find an optimal subset of the requirements with the highest value while respecting the budget. But the value of a requirement may depend on the presence or absence of other requirements in the optimal subset. Existing requirement selection methods, however, do not consider \textit{Value Dependencies}, thus increasing the risk of value loss. To address this, we have proposed \textit{Dependency-Aware Requirement Selection} (DARS) method with two main components: (i) a fuzzy-based technique for identifying and modeling value dependencies, and (ii) an \textit{Integer Programming} model that takes into account value dependencies in software requirement selection. We have further, proposed an alternative optimization model for situations where quantifying value dependencies is hard. The scalability of DARS and its effectiveness in reducing the risk of value loss are demonstrated through exhaustive simulations.


%% file: introduction.tex
\section{Introduction}
\label{sec_introduction}
Requirement selection is an essential component of \textit{Software Release Planning}~\cite{Zhang:2018:ESM:3208361.3196831,bagnall_next_2001,franch2016software,mougouei2019fuzzy,mougouei2015partial}. Traditionally, software requirement selection aims to find, for a release of software, an optimal subset of the requirements that delivers the highest economic value while respecting the project constraints such as budget~\cite{dahlstedt2005requirements}. But software requirements are known to influence the values of each other~\cite{carlshamre_industrial_2001,carlshamre_release_2002,pitangueira2015software}. As such, selecting or ignoring a requirement may impact, either positively or negatively, the values of other requirements in the optimal subset~\cite{Zhang_RIM_2013,Robinson_RIM_2003}; it is important to consider value dependencies in software requirement selection~\cite{mougouei2016factoring,mougouei2017dependency,mougouei2017modeling,mougouei2019dependency,carlshamre_industrial_2001,li_integrated_2010,zhang_investigating_2014,karlsson_improved_1997,Mougouei:2018:OHV:3236024.3264843}. However requirement dependencies, in general, and value dependencies, in particular, are \emph{fuzzy}~\cite{mougouei2020dependency,carlshamre_industrial_2001} as the strengths of those dependencies are imprecise and vary ~\cite{dahlstedt2005requirements,ngo_wicked_2008,ngo2005measuring,carlshamre_industrial_2001} from large to insignificant~\cite{wang_simulation_2012} in software projects. 

It is important thus, to consider not only the existence but the strengths of value dependencies~\cite{dahlstedt2005requirements,carlshamre_industrial_2001} and the imprecision associated with those dependencies. Existing software requirement selection methods, however, aim to find a subset of the requirements with the highest \textit{Accumulated Value} (AV)~\cite{baker_search_2006,li_integrated_2010,boschetti_lagrangian_2014,araujo2016architecture,Greer_evolutionary_2004} or \textit{Expected Value} (EV)~\cite{pitangueira2017minimizing,li2016value,li2014robust} while ignoring value dependencies among requirements. Having said that, few requirement selection methods have attempted to account for value dependencies by manually estimating the values of requirement subsets. For $n$ requirements, such methods may need up to $O(2^n)$ estimations in a worst case~\cite{van_den_akker_flexible_2005}. When this is limited to pairs of requirements, however, $O(n^2)$ estimations are needed~\cite{li_integrated_2010,sagrado_multi_objective_2013,Zhang_RIM_2013}, which is still a tedious task. Such complexity restricts the practicality of the methods that rely on manual estimations; needless to say, manual estimations are prone to inaccuracies induced by human error. Finally, estimating the values of requirement subsets does not capture the directions of value dependencies; whether a requirement $r_i$ influences $r_j$ or the other way. In other words, such methods do not distinguish among three scenarios: (i) requirement $r_i$ influences the value of requirement $r_j$ and not the other way; (ii) $r_j$ influences the value of $r_i$ and not the other way; and (iii) both $r_i$ and $r_j$ influence the values of each other but the strengths and the qualities (positive or negative) of those influences vary. 

To integrate value dependencies in software requirement selection, we have proposed a fuzzy-based optimization method that takes into account value dependencies and their imprecision when finding an optimal subset of the requirements. The proposed method, referred to as \textit{Dependency-Aware Requirement Selection} (DARS), aims to minimize the value loss caused by ignoring (selecting) requirements that have an overall positive (negative) impact on the values of the selected requirements. The DARS method comprises two major components as follows. 

(i) \textit{Identifying and modeling value dependencies}. Eells measure~\cite{eells1991probabilistic} of causal strength is used to identify and quantify value dependencies based on user preferences for software requirements. Based on fuzzy graphs~\cite{rosenfeld_fuzzygraph_1975}~\cite{kalampakas_fuzzy_2013}, a modeling technique is proposed to model value dependencies among requirements; value dependencies are modeled as fuzzy relations~\cite{carlshamre_industrial_2001,ngo_fuzzy_2005_structural,ngo2005measuring,liu_imprecise_1996}. 

(ii) \textit{Integrating value dependencies in requirement selection}. At the heart of DARS is an integer programming model, which takes the fuzzy model of value dependencies as its input and maximizes the \textit{Overall Value} (OV) of the selected requirements, where value dependencies are considered. In other words, DARS aims to minimize the value loss induced by ignoring value dependencies. We have further, contributed a complementary optimization model that allows for mitigating the risk of value loss in software projects when it is hard to quantify value dependencies. 

The effectiveness and scalability of DARS are demonstrated by carrying out simulations on the requirements of a real-world software project. Our initial results show that (a) compared to existing requirement selection methods, DARS provides a higher overall value by reducing the value loss caused by ignoring (selecting) requirements with positive (negative) influences on the values of the selected requirements, (b) maximizing the accumulated value of a requirement subset conflicts with maximizing its overall value -- where value dependencies are considered, and (c) DARS is scalable to software with a large number of requirements as well as different levels of value dependencies and precedence dependencies among those requirements. 

%% file: related.tex
\section{Related Work}
\label{sec_related}

Value is a broad notion which ranges from monetary value to other aspects of human values such as social values, fairness and so on~\cite{mougouei2018operationalizing,perera2019study,perera2019towards,hussain2018integrating}. This paper focuses on value in its economic sense. Value dependencies among software requirements~\cite{Zhang_RIM_2013,brasil_multiobjective_2012,Robinson_RIM_2003,dahlstedt2005requirements} can be explicit or implicit and vary in strength (weak and strong) and quality (positive or negative). Also, \textit{Precedence Dependencies}~\cite{dahlstedt2005requirements}, e.g. requirement $r_i$ requires (conflicts-with) $r_j$,  also have value implications; $r_i$ cannot give any value if $r_j$ is ignored (selected). As such, requirement selection methods need to consider qualities and strengths of explicit and implicit value dependencies as well as the value implications of predence dependencies among software requirements. 



\begin{table}[!htb]
	\setlength\arrayrulewidth{1.5pt}
	\caption{Considering aspects of value dependencies.}
	\label{table_lr}
	\centering
	\input{table_lr4b}
\end{table}


The first category of requirement selection methods (Table~\ref{table_lr}), i.e. \textit{Binary Knapsack} (BK) methods exploit the original formulation of binary knapsack problem~\cite{harman_exact_2014,carlshamre_industrial_2001} as given by (\ref{eq_pcbk})-(\ref{eq_pcbk_c2}). Let~$R=\{r_1,...,r_n\}$ be a set of identified requirements, where $\forall r_i\in R$ ($1 \leq i \leq n$), $v_i$ and $c_i$ in (\ref{eq_pcbk})-(\ref{eq_pcbk_c2}) denote the value and the cost of $r_i$ respectively. Also, $b$ in (\ref{eq_pcbk_c1}) denotes the available budget. A decision variable $x_i$ specifies whether requirement $r_i$ is selected ($x_i=1$) or not ($x_i=0$). The objective of \textit{BK} methods as given by (\ref{eq_pcbk}) is to find a subset of $R$ that maximizes the accumulated (or the expected) value of the selected requirements $(\sum_{i=1}^{n} v_i  x_i)$ while entirely ignoring value dependencies as well as precedence dependencies among the requirements. 

\begin{align}
\label{eq_pcbk}
& \text{Maximize} \sum_{i=1}^{n} v_i x_i \\
\label{eq_pcbk_c1}
& \text{Subject to} \sum_{i=1}^{n} c_i  x_i \leq b \\
\label{eq_pcbk_c2}
& x_i \in \{0,1\},\quad i = 1,...,n
\end{align}


The second category, i.e. \textit{Precedence-Constrained Binary Knapsack} (PCBK) methods (Table~\ref{table_lr}), comprises the majority of existing requirement selection methods, and enhances the BK methods by considering precedence dependencies and their value implications by adding (\ref{eq_pcbk_c3}) to the optimization model. A positive (negative) dependency from a requirement $r_j$ to $r_k$ is denoted by $x_j\le x_k$ ($x_j\le 1-x_k$) in (\ref{eq_pcbk_c3}). Also, decision variable $x_i$ denotes whether a requirement $r_i$ is selected ($x_i=1$) or not. 

\begin{align}
\label{eq_pcbk_c3}
&\begin{cases}
x_j \le x_k  & \text{$r_j$ positively depends on $r_k$} \\
x_j \le 1-x_k& \text{$r_j$ negatively depends on $r_k$}, \\
\end{cases}
\\ \nonumber 
& \hspace{0.5cm}j\neq k= 1,...,n
\end{align}


The third category, i.e. \textit{Increase-Decrease} methods (Table~\ref{table_lr}), consider value dependencies by estimating the amount of the increased (decreased) values induced by selecting different subsets of requirements. The optimization model of an increases-Decreases method proposed by Akker \textit{et al.}~\cite{van_den_akker_flexible_2005} is given in (\ref{eq_pcbk-CS})-(\ref{eq_pcbk-CS_c3}). For a subset $ s_j \in S:\{s_1,...,s_m\}, m\leq2^n$, with $n_j$ requirements, the difference between the estimated value ($w_j$) and the accumulated value ($\sum_{r_k\in s_j}v_k$) of $s_j$ is considered in requirement selection. $y_j$ in (\ref{eq_pcbk-CS}) specifies whether a subset $s_j$ is slected ($y_j=1$) or not ($y_j=0$). Also, constraint~(\ref{eq_pcbk-CS_c2}) ensures that $y_j=1$ only if $\forall r_k \in s_j, x_k=1$.

\begin{align}
\label{eq_pcbk-CS}
&\text{Maximize} \sum_{i=1}^{n} v_i x_i + \sum_{j=1}^{m} (w_j-\sum_{r_k\in s_j}v_k)\text{ } y_j\\
\label{eq_pcbk-CS_c1}
&\text{Subject to }  n_j y_j \leq \sum_{r_k \in s_j} x_k \\
\label{eq_pcbk-CS_c2}
&\sum_{i=1}^{n} c_i  x_i \leq b \\
\label{eq_pcbk-CS_c3}
& x_i,y_j \in \{0,1\},\quad i = 1,...,n,\quad j = 1,...,m
\end{align}

Relying on manual estimations, Increase-Decrease methods are complex and prone to human error~\cite{mougouei2016factoring}. For $n$ requirements, $O(2^n)$ estimations may be needed in worst case. This can be reduced to $O(n^2)$ when manual estimations are constrained to pairs (subsets of size 2) only~\cite{li_integrated_2010,Zhang_RIM_2013,sagrado_multi_objective_2013}. That is still a tedious task; consider $10000$ manual estimations may be needed for $100$ requirements. Finally, manual estimations, do not capture the directions of value dependencies: (a) $r_i$ influences the value of $r_j$; (b) $r_j$ influences the value of $r_i$; and (c) $r_i$ and $r_j$ influence the values of each other but the strengths and the qualities (positive or negative) of those influences vary.

%% file: table_lr4b.tex
\large
\resizebox {0.4\textwidth }{!}{
	\begin{tabular}{|l|l|}
		\toprule[1.5 pt]
		\textbf{\cellcolor{header}\textcolor{white}{Category}} &
		\textbf{\cellcolor{header}\textcolor{white}{Requirement Selection Methods}} 
		\bigstrut\\
		\hline
		BK &
		\begin{tabular}[t]{@{}c@{}} \cite{karlsson_optimizing_1997,jung_optimizing_1998,zhang_multi_objective_2007,baker_search_2006,finkelstein2009search,zhang2011comparing,del2010ant,kumari2012software} \phantom{sssssssssssss} \end{tabular} 
		\bigstrut\\
		\hline
		\multirow{4}[9]{*}{PCBK} &
		\begin{tabular}[t]{@{}l@{}} \cite{veerapen2015integer,brasil_multiobjective_2012,sagrado_multi_objective_2013,bagnall_next_2001,greer_software_2004,boschetti_lagrangian_2014,ruhe_quantitative_2003,van2011quantitative,zhang2010search,tonella2010using,freitas2011software}\\ \cite{colares_new_2009,saliu2007bi,saliu2005supporting,jiang2010hybrid,van2005determination,ngo2009optimized,chen2013ant,del2011requirements,araujo2016architecture,colares2009new,pitangueira2016risk}\\\cite{van2008software,ngo_wicked_2008,tonella2013interactive,xuan2012solving,saliu2005software,pitangueira2017minimizing,li2016value,li2014robust}\end{tabular} 
		\bigstrut\\
		\hline
		\multicolumn{1}{|l|}{\multirow{2}[4]{*}{Increase-Decrease}} &
		\cite{li_integrated_2010,sagrado_multi_objective_2013,Zhang_RIM_2013} (subsets of size 2) 
		\bigstrut\\
		\cline{2-2}\multicolumn{1}{|l|}{} &
		\cite{van_den_akker_flexible_2005} (subsets of any size) 
		\bigstrut\\
		\hline
	\end{tabular}%
}

%% file: identificationANDmodeling.tex
\section{Modeling and Identification}
\label{dars_identificationAndModeling}

This section presents a technique for identifying explicit value dependencies among software requirements. Explicit dependencies will then, be used in Section~\ref{ch_dars_modeling} to infer implicit dependencies and compute the overall influences of requirements on the values of each other. 

%% file: identification.tex
\subsection{Identifying Explicit Value Dependencies}
\label{ch_dars_identification}
It is widely recognized that the values of software requirements (features) are determined by users preferences~\cite{zhang2019software,mougouei2017modeling}. But the preferences of users for certain requirements may change in the presence or absence of other requirements. Such causal relations among requirements thus can be exploited for detecting value dependencies; measures of causal strength~\cite{Halpern01062015,pearl2009causality,janzing2013quantifying} can be used to quantify value dependencies. In this regard, conventional market research techniques such as surveys and mining user reviews, e.g. mining Google Play and App Store, can be used to collect user preferences for software requirements~\cite{leung2011probabilistic,holland2003preference,sayyad2013value,villarroel2016release}. We have proposed using Eells measure~\cite{eells1991probabilistic} of causal strength -- due to its effectiveness in capturing both positive and negative causation~\cite{sprenger2016foundations} -- to quantify the strengths and qualities of explicit value dependencies among requirements as given by (\ref{Eq_ch_dars_Eells}). The sign (magnitude) of $\eta_{i,j}$ gives the quality (strength) of a value dependency from requirement $r_i$ to $r_j$, where selecting (ignoring) $r_j$ may influence, positively or negatively, the value of $r_i$. 

\begin{align}
\label{Eq_ch_dars_Eells}
& \eta_{i,j}= p(r_i|r_j) - p(r_i|\bar{r_j}) ,\phantom{s}\eta_{i,j} \in [-1,1]
\end{align}

For a pair of requirements ($r_i,r_j$), Eells measure captures both positive and negative value dependencies from $r_i$ to $r_j$ by deducting the conditional probability $p(r_i|\bar{r_j})$ from $p(r_i|r_j)$, where $p(r_i|\bar{r_j})$ and $p(r_i|r_j)$ denote strengths of positive and negative causal relations from $r_i$ to $r_j$ respectively, that is selecting $r_i$ may increase or decrease the value of $r_j$.

\begin{figure}[!htb]
	\begin{center}
		\subfigure[$P_{4 \times 4}$]{%
			\label{fig_ch_dars_eta_p}
			\includegraphics[scale=0.3]{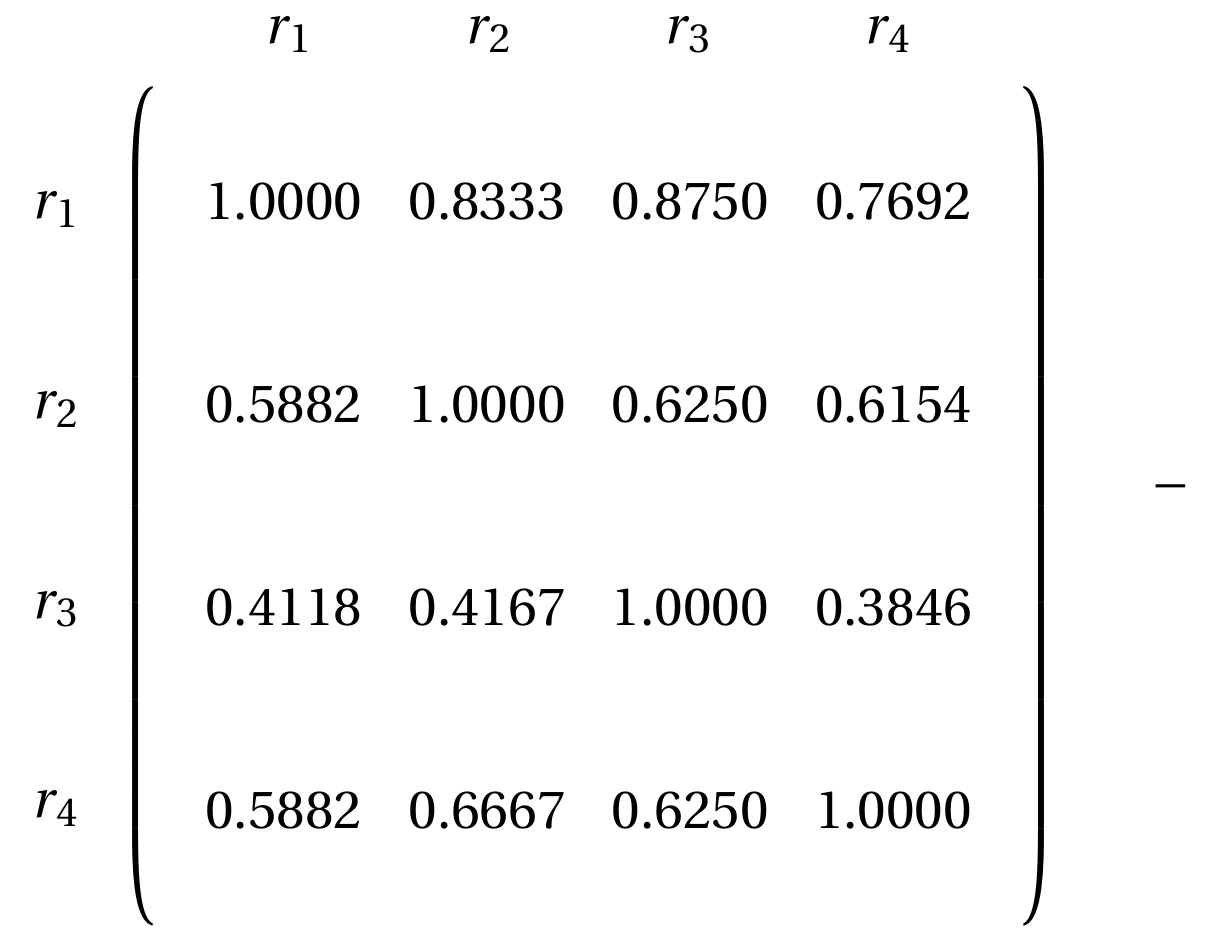}
		}
		\subfigure[$\bar{P}_{4 \times 4}$]{%
			\label{fig_ch_dars_eta_p_bar}
			\includegraphics[scale=0.3]{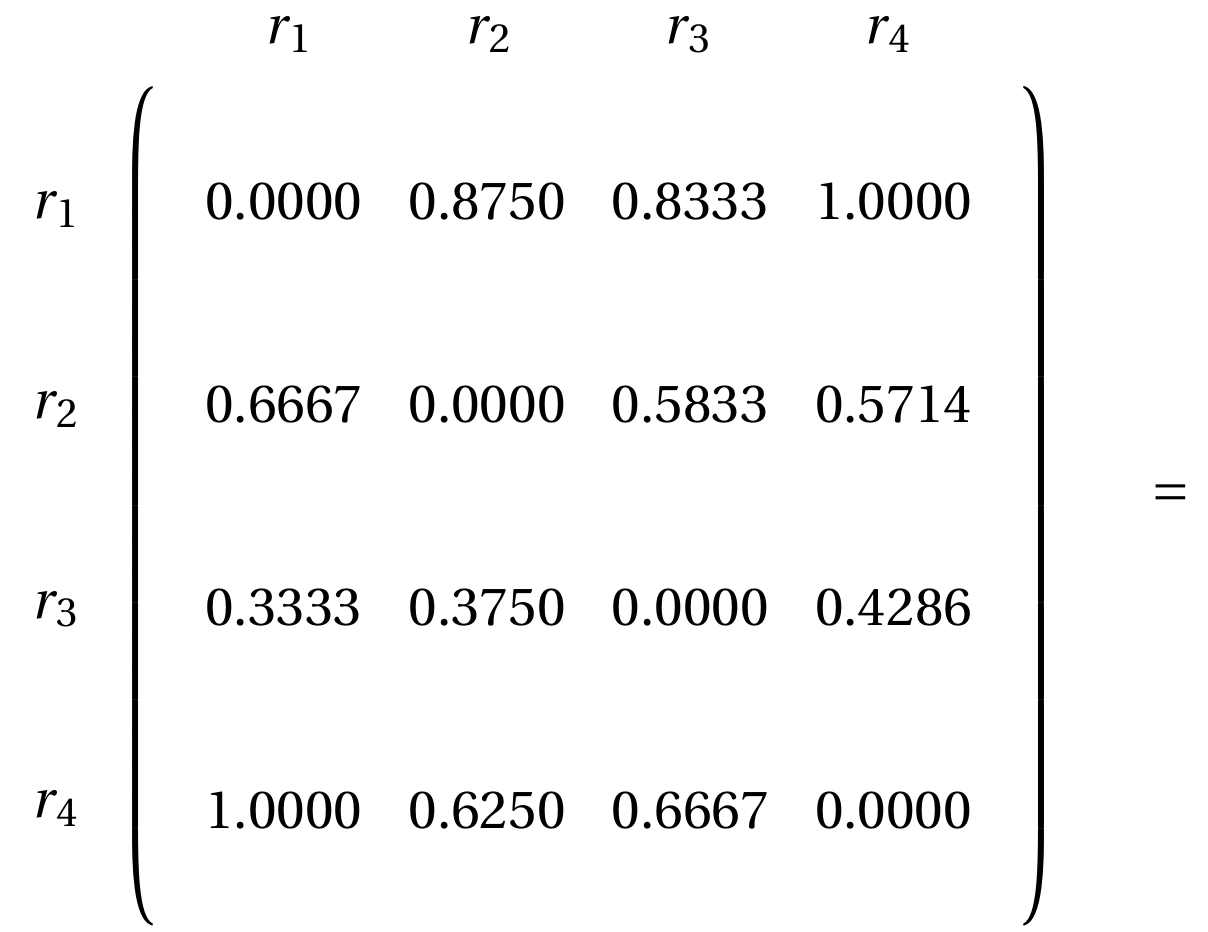}
		}
		\subfigure[$\bm{\eta}_{4 \times 4}$]{%
			\label{fig_ch_dars_eta_eta}
			\includegraphics[scale=0.3]{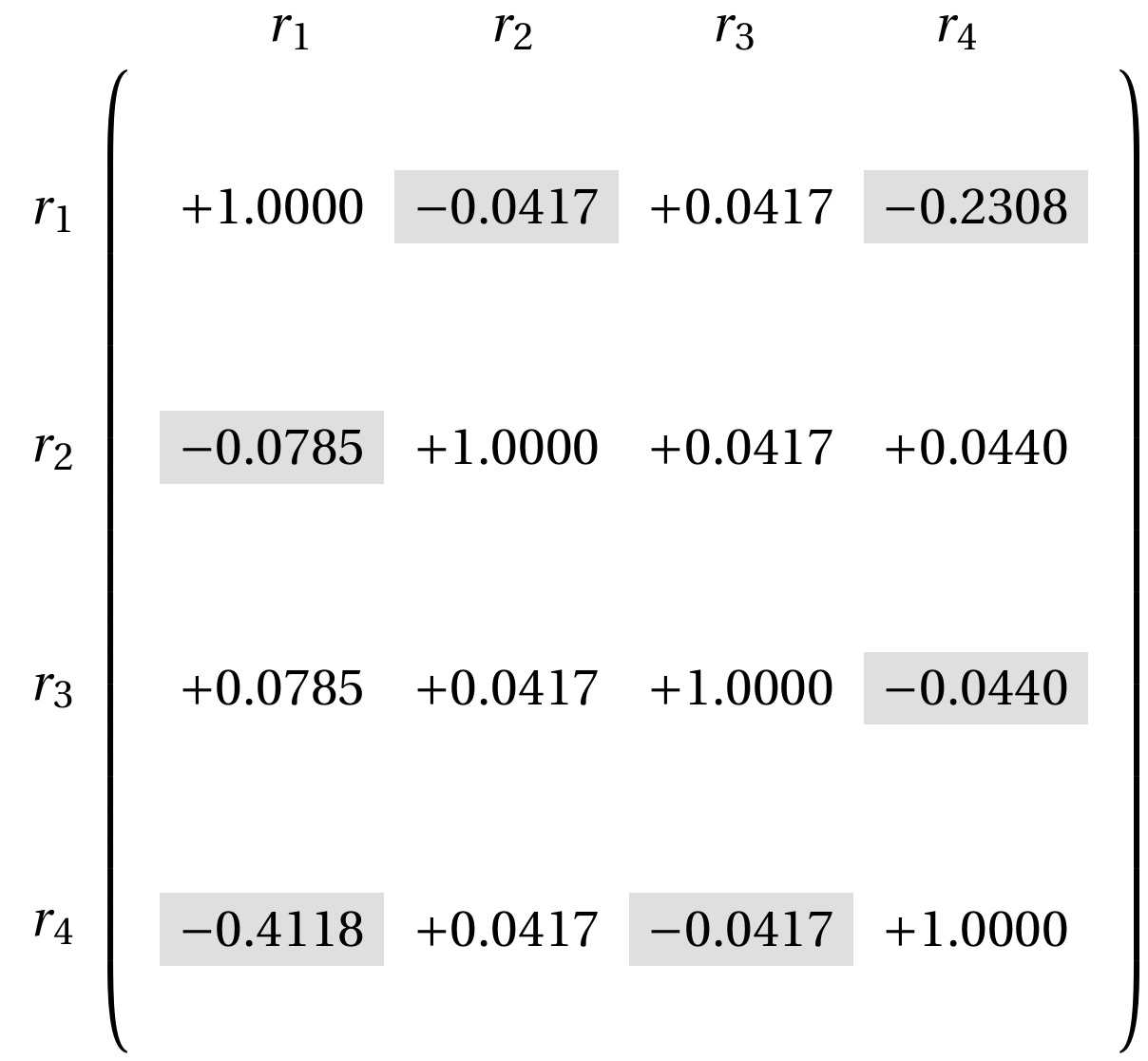}
		}
	\end{center}
	\vspace{-0.25cm}
	\caption{%
		Computing Eells measure for Figure~\ref{fig_ch_dars_pm}.
	}%
	\label{fig_ch_dars_eta}
\end{figure}

\begin{figure}
	\begin{center}
		\includegraphics[scale=0.7]{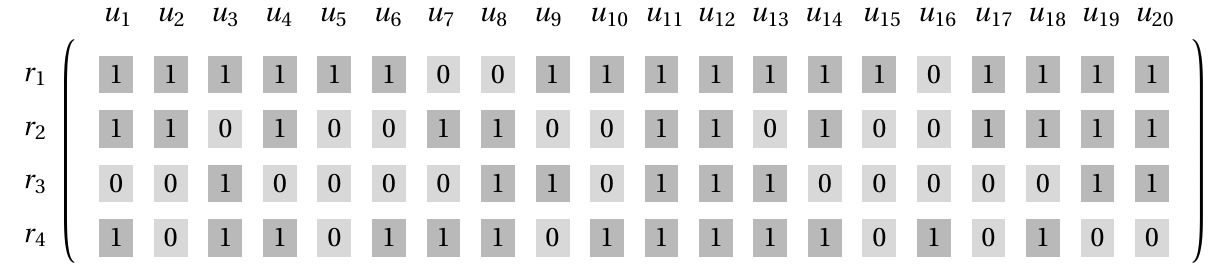}
	\end{center}
	\vspace{-0.25cm}
	\caption{Preference matrix $M_{4\times 20}$.}
	\label{fig_ch_dars_pm}
\end{figure}

\begin{exmp}
	\label{ex_ultimate_strength}
	Matrix $M_{4\times 20}$ in Figure~\ref{fig_ch_dars_pm} shows the preferences of $20$ users ($u_1,...,u_{20}$) for $4$ requirements ($r_1,...,r_4$). Each element $m_{i,j}$ specifies whether requirement $r_j$ has been preferred by user $u_i$ ($m_{i,j}=1$) or not ($m_{i,j}=0$). Matrix $P_{4\times4}$ (Figure~\ref{fig_ch_dars_eta_p}) and Matrix $\bar{P}_{4\times4}$ (Figure~\ref{fig_ch_dars_eta_p_bar}) show the strengths of causal relations among user preferences for requirements in $M_{4\times 8}$. For a pair of requirements $r_i$ and $r_j$ with $i \neq j$, an off-diagonal element $p_{i,j}$ ($\bar{p}_{i,j}$) of matrix $P_{4\times4}$ ($\bar{P}_{4\times4}$) denotes the strength of a positive (negative) causal relation from $r_i$ to $r_j$. For diagonal elements of $P_{4\times4}$ ($\bar{P}_{4\times4}$) on the other hand, we have $p_{i,i}=p(r_i|r_i)=1$ ($\bar{p}_{i,i}=p(r_i|\bar{r}_i)=0$). Hence, subtracting each element $\bar{p}_{i,j}$ from its corresponding element $p_{i,j}$, where $i \neq j$, gives Eells causal strength $\eta_{i,j}$ for the value dependency from $r_i$ to $r_j$. Diagonal elements, however, may be ignored as they denote self-causation; $p(r_i|r_i)=1$.
\end{exmp}

\begin{figure}[!htb]
	\begin{center}
		\subfigure[$$]{%
			\label{fig_ch_dars_membership_1}
			\includegraphics[scale=0.8]{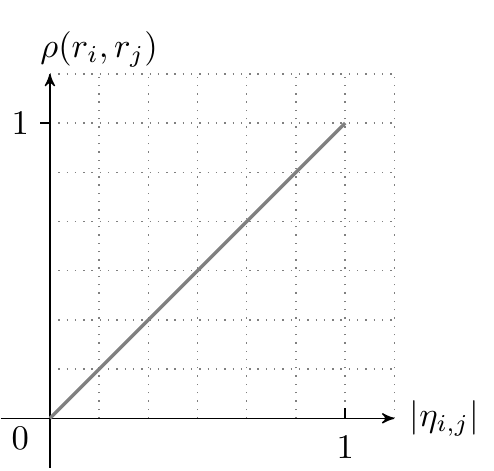}
		}
		\subfigure[$$]{%
			\label{fig_ch_dars_membership_2}
			\includegraphics[scale=0.8]{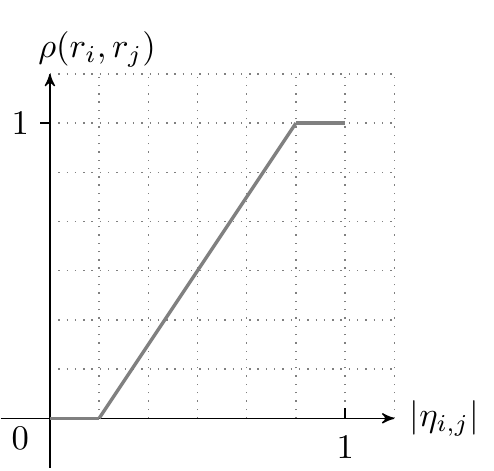}
		}
	\end{center}
	\captionsetup{margin=0ex}
	\vspace{-0.25cm}
	\caption{%
		Sample membership functions for value dependencies.
	}%
	\label{fig_ch_dars_membership}
\end{figure}

The strength of an explicit dependency from a requirement $r_i$ to $r_j$ is computed by (\ref{Eq_ch_dars_strengthMeasure}), which gives a mapping from Eells measure $\eta_{i,j}$ to $\rho: R\times R\rightarrow [0,1]$ (Figure~\ref{fig_ch_dars_membership}). As given by (\ref{Eq_ch_dars_qualityMeasure}), $\eta_{i,j}>0$ indicates that the strength of the positive causal relation from $r_i$ to $r_j$ is greater than the strength of its corresponding negative relation: $p(r_i|r_j) > p(r_i|\neg r_j)$; the quality of $(r_i,r_j)$ is positive ($\sigma(r_i,r_j)=+$). Similarly, $\eta_{i,j}<0$ indicates $p(r_i|\neg r_j) > p(r_i|r_j) \rightarrow \sigma(r_i,r_j)=-$. Also, $p(r_i|r_j) - p(r_i|\neg r_j)=0$ specifies that the quality of the zero-strength dependency $(r_i,r_j)$ is non-specified ($\sigma(r_i,r_j)=\pm$).

\begin{align}
\label{Eq_ch_dars_strengthMeasure}
& \rho(r_i,r_j)= |\eta_{i,j}|\\[10pt]
\label{Eq_ch_dars_qualityMeasure}
& \sigma(r_i,r_j) =  \begin{cases}
+ & \text{if }\phantom{s}  \eta_{i,j} > 0 \\
- & \text{if }\phantom{s}  \eta_{i,j} < 0 \\
\pm & \text{if }\phantom{s} \eta_{i,j} = 0 \\
\end{cases}
\end{align}

Different membership functions may be adopted to account for the preferences of decision makers, e.g. the membership function in Figure~\ref{fig_ch_dars_membership_1} may be used to ignore ``very weak'' value dependencies while ``very strong'' dependencies are considered as full strength relations, $\rho(r_i,r_j)=1$. 

%% file: modeling.tex
\subsection{Modeling Value Dependencies}
\label{ch_dars_modeling}

Fuzzy logic has been widely adopted in decision making~\cite{rosenfeld_fuzzygraph_1975} to capture the imprecision of real-world problems~\cite{mougouei2013fuzzy,mougouei2013goal,Mathew_strong_2013,mougouei2012evaluating,mougouei2012measuring,mougouei2012goal}. In Software Engineering, fuzzy graphs have demonstrated useful in modeling the imprecision of dependency relations in software~\cite{mougouei2018mathematical,ngo_fuzzy_2005_structural,ngo2005measuring}. Ngo-The \textit{et al.} used fuzzy graphs for modeling dependency satisfaction in software release planning~\cite{ngo_fuzzy_2005_structural} and capturing the imprecision of coupling dependencies among software requirements~\cite{ngo2005measuring}. Also, Wang \textit{et al.}~\cite{wang_simulation_2012} exploited fuzzy logic to capture the strengths of dependency relations among requirements. In this section, we discuss modeling value dependencies among requirements by fuzzy graphs. We further use the algebraic structure of fuzzy graphs to compute the overall influences of requirements on the values of each other.   

\subsubsection{Value Dependency Graphs}
\label{ch_dars_modeling_vdg}

To account for the imprecision of value dependencies, we have introduced \textit{Value Dependency Graphs} (VDGs) based on fuzzy graphs for modeling value dependencies and their characteristics. We have specially modified the classical definition of fuzzy graphs to consider not only the strength but also the quality (positive or negative) of value dependencies as given by Definition~\ref{def_vdg}. 

\begin{figure}[!htbp]
	\begin{center}
		\includegraphics[scale=0.75]{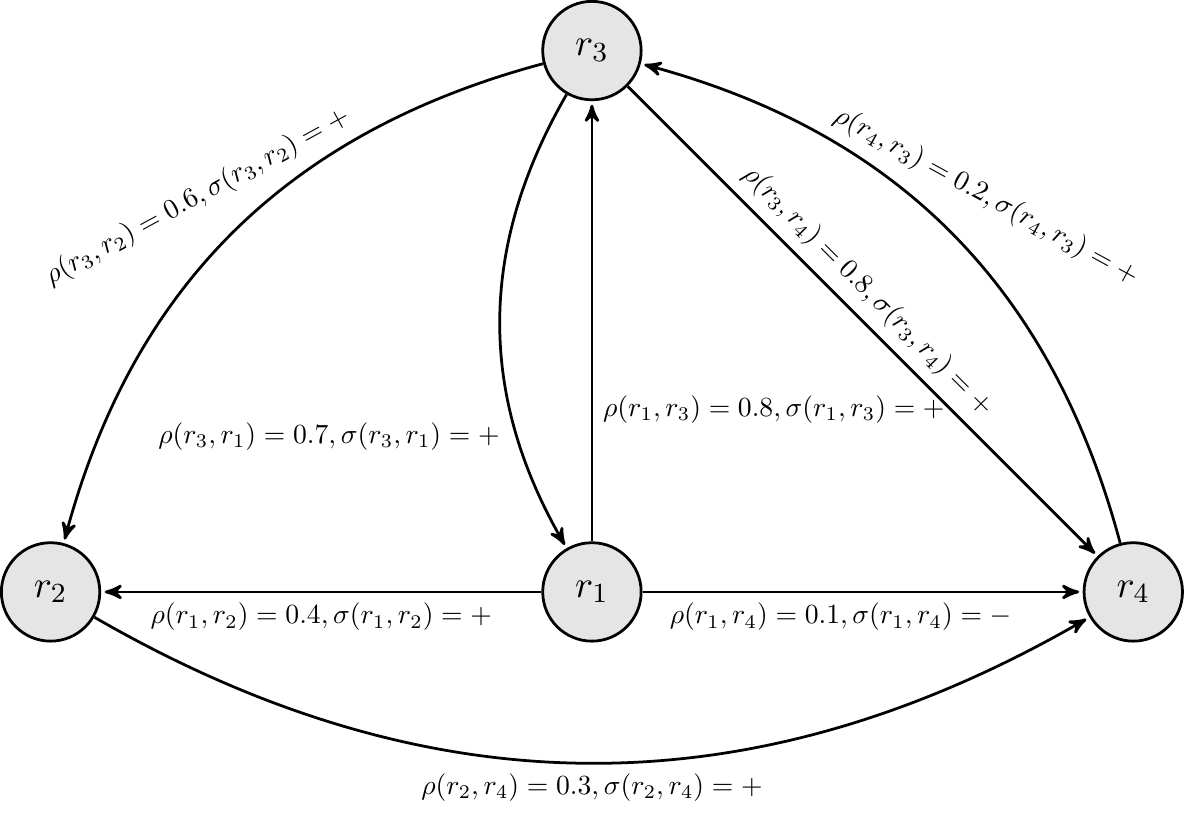}
	\end{center}
	\caption{%
		A sample value dependency graph.}%
	\label{fig_ch_dars_ex_vdg}
\end{figure}

\begin{mydef}
	\label{def_vdg}
	\textit{The Value Dependency Graph} (VDG) is a signed directed fuzzy graph~\cite{Wasserman1994} $G=(R,\sigma,\rho)$ where, requirements $R:\{r_1,...,r_n\}$ constitutes the graph nodes. Also, the qualitative function $\sigma(r_i,r_j) \rightarrow \{+,-,\pm\}$ and the membership function $\rho: (r_i,r_j)\rightarrow [0,1]$ denote the quality and the strength of the explicit value dependency (edge of the graph) from $r_i$ to $r_j$ receptively. Moreover, $\rho(r_i,r_j)=0$ denotes the absence of any explicit value dependency from $r_i$ to $r_j$. In that case we have $\sigma(r_i,r_j)=\pm$, where $\pm$ denotes the quality of the dependency is non-specified; in Figure~\ref{fig_ch_dars_ex_vdg}, $\sigma(r_1,r_2)=+$ and $\rho(r_1,r_2)=0.4$ specifies a positive value dependency from $r_1$ to $r_2$ with strength $ 0.4 $. 
\end{mydef}

\subsubsection{Value Dependencies in VDGs}
\label{ch_dars_modeling_vdg}

In Section~\ref{ch_dars_identification} we introduced an automated technique for the identification of explicit value dependencies and their characteristics (quality and strength) from user preferences. Definition~\ref{def_vdg_valuedepndencies} provides a more comprehensive definition of value dependencies that includes both explicit and implicit value dependencies among the requirements of software based on the algebraic structure of fuzzy graphs. 

\begin{mydef}
	\label{def_vdg_valuedepndencies}
	\textit{Value Dependencies}. 
	A value dependency in a value dependency graph $G=(R,\sigma,\rho)$ is defined as a sequence of requirements $d_i:\big(r(0),...,r(k)\big)$ such that $\forall r(j) \in d_i$, $1 \leq j \leq k$ we have $\rho\big(r(j-1),r(j)\big) \neq 0$. $j\geq 0$ is the sequence of the $j^{th}$ requirement (node) denoted as $r(j)$ on the dependency path. A consecutive pair $\big(r(j-1),r(j)\big)$ specifies an explicit value dependency. 
\end{mydef}

\begin{align}
\label{Eq_ch_dars_vdg_strength}
&\forall d_i:\big(r(0),...,r(k)\big): \rho(d_i) = \bigwedge_{j=1}^{k}\text{ }\rho\big(r(j-1),r(j)\big) \\
\label{Eq_ch_dars_vdg_quality}
&\forall d_i:\big(r(0),...,r(k)\big): \sigma(d_i) = \prod_{j=1}^{k}\text{ }\sigma\big(r(j-1),r(j)\big)
\end{align}

Equation (\ref{Eq_ch_dars_vdg_strength}) computes the strength of a value dependency $d_i:\big(r(0),...,r(k)\big)$ by finding the strength of the weakest of $k$ explicit dependencies on $d_i$.  $\wedge$ denotes fuzzy AND operation~\cite{zadeh_fuzzysets_1965}. The quality (positive or negative) of a value dependency $d_i:\big(r(0),...,r(k)\big)$ is calculated by qualitative serial inference~\cite{de1984qualitative,wellman1990formulation,kusiak_1995_dependency} as given by (\ref{Eq_ch_dars_vdg_quality}) and Table~\ref{table_ch_dars_inference}. Inferences in Table~\ref{table_ch_dars_inference} are informally proved by Wellman~\cite{wellman1990formulation} and Kleer~\cite{de1984qualitative}. 

\begin{table}[!htb]
	\caption{Qualitative serial inference in VDGs.}
	\label{table_ch_dars_inference}
	\centering
	\input{table_inference}

\end{table}


Let $D=\{d_1,d_2,..., d_m\}$ be the set of all value dependencies from $r_i \in R$ to $r_j \in R$ in a VDG $G=(R,\sigma,\rho)$, where positive and negative dependencies can simultaneously exist from $r_i$ to $r_j$. The strength of all positive dependencies from $r_i$ to $r_j$, denoted by $\rho^{+\infty}(r_i,r_j)$, is calculated by (\ref{Eq_ch_dars_ultimate_strength_positive}), that is to find the strongest positive dependency~\cite{rosenfeld_fuzzygraph_1975} from $r_i$ to $r_j$. Fuzzy operators $\wedge$ and $\vee$ denote Zadeh's~\cite{zadeh_fuzzysets_1965} AND and OR operations respectively. Analogously, the strength of all negative value dependencies from $r_i$ to $r_j$ is denoted by $\rho^{-\infty}(r_i,r_j)$ and calculated by (\ref{Eq_ch_dars_ultimate_strength_negative}).

\begin{align}
\label{Eq_ch_dars_ultimate_strength_positive}
&\rho^{+\infty}(r_i,r_j) = \bigvee_{d_m\in D, \sigma(d_m)=+} \text{ } \rho(d_m) \\[2pt]
\label{Eq_ch_dars_ultimate_strength_negative}
&\rho^{-\infty}(r_i,r_j) = \bigvee_{d_m\in D, \sigma(d_i)=-} \text{ } \rho(d_m) \\
\label{Eq_ch_dars_influence}
&I_{i,j} = \rho^{+\infty}(r_i,r_j)-\rho^{-\infty}(r_i,r_j) 
\end{align}

A brute-force approach to computing $\rho^{+\infty}(r_i,r_j)$ or $\rho^{-\infty}(r_i,r_j)$ needs to calculate the strengths of all paths from $r_i$ to $r_j$, which is of $O(n!)$ for $n$ requirements. To reduce this complexity, we have devised a modified version of Floyd-Warshall~\cite{floyd_1962} algorithm (Algorithm~\ref{alg_ch_dars_strength}), that computes $\rho^{+\infty}(r_i,r_j)$ and $\rho^{-\infty}(r_i,r_j)$ for all pairs of requirements $(r_i,r_j),\text{ }r_i,r_j \in R:\{r_1,...,r_n\}$ in $O(n^3)$. For each pair of requirements $(r_i,r_j)$ in a VDG $G=(R,\sigma,\rho)$, lines $18$ to $35$ of Algorithm~\ref{alg_ch_dars_strength} find the strength of all positive (negative) value dependencies from $r_i$ to $r_j$. The overall strength of all positive and negative value dependencies from $r_i$ to $r_j$ is referred to as the \textit{Influence} of $r_j$ on the value of $r_i$ and denoted by $I_{i,j}$. $I_{i,j}$, as in (\ref{Eq_ch_dars_influence}), is computed by deducting the strength of all negative value dependencies from $r_i$ to $r_j$ ($\rho^{-\infty}(r_i,r_j)$) from the strength of all positive value dependencies from $r_i$ to $r_j$ ($\rho^{+\infty}(r_i,r_j)$). Therefore, $I_{i,j}>0$ ($I_{i,j}<0$) states that $r_j$ positively (negatively) influences the value of $r_i$.

\begin{algorithm}
	\small
	\caption{ Strengths of value dependencies.}
	\label{alg_ch_dars_strength}
	\begin{algorithmic}[1]
		\REQUIRE VDG $G=(R,\sigma,\rho)$
		\ENSURE $\rho^{+\infty}, \rho^{-\infty}$
		\FOR{\textbf{each} $r_i \in R$}
		\FOR{\textbf{each} $r_j \in R$}
		\STATE $\rho^{+\infty}(r_i,r_j) \leftarrow \rho^{-\infty}(r_i,r_j) \leftarrow -\infty$ 
		\ENDFOR
		\ENDFOR
		\FOR{\textbf{each} $r_i \in R$}
		\STATE $\rho(r_i,r_i)^{+\infty} \leftarrow \rho(r_i,r_i)^{-\infty} \leftarrow 0$
		\ENDFOR
		\FOR{\textbf{each} $r_i \in R$}
		\FOR{\textbf{each} $r_j \in R$}
		\IF{$\sigma(r_i,r_j) = +$}
		\STATE $\rho^{+\infty}(r_i,r_j) \leftarrow \rho(r_i,r_j)$
		\ELSIF{$\sigma(r_i,r_j) = -$}
		\STATE $\rho^{-\infty}(r_i,r_j) \leftarrow \rho(r_i,r_j)$
		\ENDIF
		\ENDFOR
		\ENDFOR
		\FOR{\textbf{each} $r_k \in R$}
		\FOR{\textbf{each} $r_i \in R$}
		\FOR{\textbf{each} $r_j \in R$}
		\IF{$min\big(\rho^{+\infty}(r_i,r_k), \rho^{+\infty}(r_k,r_j)\big) > \rho^{+\infty}(r_i,r_j)$}
		\STATE $\rho^{+\infty}(r_i,r_j) \leftarrow  min(\rho^{+\infty}(r_i,r_k), \rho^{+\infty}(r_k,r_j))$
		\ENDIF
		\IF{$min\big(\rho^{-\infty}(r_i,r_k), \rho^{-\infty}(r_k,r_j)\big) > \rho^{+\infty}(r_i,r_j)$}
		\STATE $\rho^{+\infty}(r_i,r_j) \leftarrow  min(\rho^{-\infty}(r_i,r_k), \rho^{-\infty}(r_k,r_j))$
		\ENDIF
		\IF{$min\big(\rho^{+\infty}(r_i,r_k), \rho^{-\infty}(r_k,r_j)\big) > \rho^{-\infty}(r_i,r_j)$}
		\STATE $\rho^{-\infty}(r_i,r_j) \leftarrow  min(\rho^{+\infty}(r_i,r_k), \rho^{-\infty}(r_k,r_j))$
		\ENDIF
		\IF{$min\big(\rho^{-\infty}(r_i,r_k), \rho^{+\infty}(r_k,r_j)\big) > \rho^{-\infty}(r_i,r_j)$}
		\STATE $\rho^{-\infty}(r_i,r_j) \leftarrow  min(\rho^{-\infty}(r_i,r_k), \rho^{+\infty}(r_k,r_j))$
		\ENDIF
		\ENDFOR
		\ENDFOR
		\ENDFOR
	\end{algorithmic}
\end{algorithm}

%
%

\begin{exmp}
	\label{ex_ultimate_strength}
	Let $D=\{d_1:(r_1,r_2,r_4),d_2:(r_1,r_3,r_4),d_3:(r_1,r_4)\}$ be the value dependencies from $r_1$ to $r_4$ in Figure \ref{fig_ch_dars_ex_vdg}. Using (\ref{Eq_ch_dars_vdg_quality}), qualities of $d_1$ to $d_3$ are: $\sigma(d_1)=\Pi(+,+)=+$, $\sigma(d_2)=\Pi(+,+)=+$, and $\sigma(d_3)=\Pi(-)=-$. Strengths are calculated by (\ref{Eq_ch_dars_vdg_strength}) as: $\rho(d_1)=\wedge\big(\rho(r_1,r_2),\rho(r_2,r_4)\big)=min(0.4,0.3)$, $\rho(d_2)=\wedge\big(\rho(r_1,r_3),\rho(r_3,r_4)\big)=min(0.8,0.8)$, $\rho(d_3)=min(0.1)$. Then, using~(\ref{Eq_ch_dars_ultimate_strength_positive}) and~(\ref{Eq_ch_dars_ultimate_strength_negative}, we have $\rho(r_1,r_4)^{+\infty} = \vee(\rho(d_1),\rho(d_2))=max(0.3,0.8)$ and $\rho^{-\infty}(r_1,r_4) = max(\rho(d_3))$. Therefore, $I_{1,4} = \rho(r_1,r_4)^{+\infty}-\rho(r_1,r_4)^{-\infty}=0.7$. Table~\ref{table_ch_dars_ex_overall} lists influences of requirements in the VDG of Figure~\ref{fig_ch_dars_ex_vdg} on the value of each other.
\end{exmp}
\begin{table}[!htb]
	\centering
	\caption{Overall influences computed for VDG of Figure~\ref{fig_ch_dars_ex_vdg}.}
	\label{table_ch_dars_ex_overall}
	\input{table_ex_strengths}

\end{table}
\begin{mydef}
	\label{def_frig_VDL}
	\textit{Value Dependency Level (VDL) and Negative Value Dependency Level (NVDL)}. Let $G=(R,\sigma,\rho)$ be a VDG with $R=\{r_1,...,r_n\}$, $k$ be the total number of explicit value dependencies in $G$, and $m$ be the total number of negative explicit value dependencies. Then the VDL and NVDL of $G$ are derived by (\ref{Eq_ch_dars_vdl}) and (\ref{Eq_ch_dars_nvdl}) respectively. 
\end{mydef}

\begin{align}
\label{Eq_ch_dars_vdl}
&VDL(G)=\frac{k}{\Perm{n}{2}}=\frac{k}{n (n-1)} \\
\label{Eq_ch_dars_nvdl}
&NVDL(G)=\frac{m}{k}
\end{align}

\begin{exmp}
	\label{ex_vdl}
	For the value dependency graph $G$ of Figure~\ref{fig_ch_dars_ex_vdg}, we have $n=4$, $k=8$, and $m=1$. $VDL(G)$ is derived by~(\ref{Eq_ch_dars_vdl}) as: $VDL(G)= \frac{8}{4\times 3} = \frac{8}{12} \approxeq 0.67$. Also we have from Equation~(\ref{Eq_ch_dars_nvdl}), $NVDL(G)=\frac{1}{8}=0.125$.
\end{exmp}

%% file: table_inference.tex
\resizebox {0.4\textwidth }{!}{
	\begin{tabular}{cc|ccc}
		\toprule[1.5pt]
		\multicolumn{2}{r|}{\multirow{2}[1]{*}{ $\sigma\big(r(j-1),r(j),r(j+1)\big)$}} &
		\multicolumn{3}{c}{$\sigma\big(r(j),r(j+1)\big)$}
		\\
		\multicolumn{2}{r|}{} &
		$+$ &
		$-$ &
		$\pm$
		\bigstrut[b]\\
		\hline
		\multicolumn{1}{c}{\multirow{3}[1]{*}{$\sigma\big(r(j-1),r(j)\big)$}} &
		$+$ &
		$+$ &
		$-$ &
		$\pm$
		\bigstrut[t]\\
		\multicolumn{1}{c}{} &
		$-$ &
		$-$ &
		$+$ &
		$\pm$
		\\
		\multicolumn{1}{c}{} &
		$\pm$ &
		$\pm$ &
		$\pm$ &
		$\pm$
		\\
    \bottomrule[1.5pt]
	\end{tabular}%
	}

%% file: table_ex_strengths.tex
\footnotesize
\resizebox {0.2\textwidth }{!}{
	\begin{tabular}{lcccc}
		\toprule[1.5pt]
		\textbf{\cellcolor{white}\textcolor{black}{$I_{i,j}$}}&
		\textbf{\cellcolor{white}\textcolor{black}{$r_1$}}&
		\textbf{\cellcolor{white}\textcolor{black}{$r_2$}}&
		\textbf{\cellcolor{white}\textcolor{black}{$r_3$}}&
		\textbf{\cellcolor{white}\textcolor{black}{$r_4$}}
		\\ \midrule
		\textbf{$r_1$}\unboldmath{} &
		$0.0$ &
		$0.5$ &
		$0.7$ &
		$0.7$
		\\
		\textbf{$r_2$}\unboldmath{} &
		$0.2$ &
		$0.0$ &
		$0.2$ &
		$0.3$
		\\
		\textbf{$r_3$}\unboldmath{} &
		$0.6$ &
		$0.5$ &
		$0.0$ &
		$0.7$
		\\
		\textbf{$r_4$}\unboldmath{} &
		$0.2$ &
		$0.2$ &
		$0.2$ &
		$0.0$
		\\ \bottomrule[1.5pt]
	\end{tabular}
}

%% file: selection.tex
\section{Integrating Value Dependencies}
\label{ch_dars_selection}

%% file: selection_ov.tex
\subsection{Overall Value of a Requirement Subset}
\label{ch_dars_selection_ov}

This section details our proposed measure of value, i.e. overall value (OV), for the economic worth of a requirement subset. OV takes into account user preferences for the selected requirements as well as the impacts of value dependencies on the values of requirements. Value dependencies, as explained in Section~\ref{ch_dars_identification}, are identified based on causal relations among user preferences. Section~\ref{ch_dars_identification} presented an automated technique for the identification of value dependencies among requirements; Algorithm~\ref{alg_ch_dars_strength} was used to infer implicit value dependencies and compute the influences of requirements on the values of each other based on the algebraic structure of fuzzy graphs. To compute the overall values of the selected requirements, (\ref{eq_ch_dars_penalty})-(\ref{eq_ch_dars_penalty_c1}) give the penalty of ignoring (selecting) requirements with positive (negative) influence on the values of the selected requirements. $\theta_{i}$ in this equation denotes the penalty for a requirement $r_i$, $n$ denotes the number of requirements, and $x_j$ specifies whether requirement $r_j$ is selected ($x_j=1$) or not ($x_j=0$). Also, $I_{i,j}$, as in (\ref{Eq_ch_dars_influence}), gives the positive or negative influence of $r_j$ on the value of $r_i$.  

\begin{align}
\label{eq_ch_dars_penalty}
\nonumber
\theta_{i}= &\displaystyle \bigvee_{j=1}^{n} \bigg(\frac{x_j\big(\lvert I_{i,j} \rvert-I_{i,j}\big) + (1-x_j)\big(\lvert I_{i,j}\rvert+I_{i,j}\big)}{2}\bigg)=\\ 
&\displaystyle \bigvee_{j=1}^{n} \bigg(\frac{\lvert I_{i,j} \rvert + (1-2x_j)I_{i,j}}{2}\bigg),i \neq j =1,...,n \\
\label{eq_ch_dars_penalty_c1}
& x_j \in\{0,1\},  j=1,...,n 
\end{align}

We made use of the algebraic structure of fuzzy graphs for computing the influences of requirements on the values of each other as explained in Section~\ref{ch_dars_modeling}. Accordingly, $\theta_i$ is computed using the fuzzy OR operator which is to take supremum over the strengths of all ignored positive dependencies and selected negative dependencies of $r_i$ in its corresponding value dependency graph. Overall values of the selected requirements thus can be computed by (\ref{eq_ch_dars_vprime}), where $v_i^{\prime}$ denotes the overall value of a requirement $r_i$, $E(v_i)$ specifies the expected value of $r_i$, and $\theta_{i}$ denotes the penalty of ignoring (selecting) positive (negative) value dependencies of $r_i$. Equation (\ref{eq_ch_dars_ov}) derives the overall value of $n$ requirements, where cost and expected value of a requirements $r_i$ are denoted by $c_i$ and $E(v_i)$ respectively. Decision variable $x_i$ specifies whether $r_i$ is selected ($x_i=1$) or not ($x_i=0$). $E(V_i)$ is computed by (\ref{eq_ch_dars_expected}), where $v_i$ denotes the estimated value of $r_i$. Also $p(r_i)$ ($p(\bar{r_i})$) specify the probability that users select (ignore) $r_i$. 

\begin{align}
\label{eq_ch_dars_expected}
&E(v_i) = p(r_i)\times v_i + p(\bar{r_i})\times 0= p(r_i)\times v_i
\end{align}

For a requirement $r_i$, $\theta_i$ specifies the penalty of ignoring (selecting) requirements with positive (negative) influence on the expected value of $r_i$. $\theta_iv_i$ in (\ref{eq_ch_dars_ov}) therefore, gives the value loss for a requirement $r_i$ as a result of ignoring (selecting) requirements that positively (negatively) impact user preferences for $r_i$ and consequently its expected value. 

\begin{align}
\label{eq_ch_dars_vprime}
& v_i^{\prime} = (1-\theta_i)E(v_i)\\
\label{eq_ch_dars_ov}
&OV = \sum_{i=1}^{n} x_i (1-\theta_i)E(v_i), \textit{ } x_i \in \{0,1\}
\end{align}

\begin{exmp}
	\label{ex_ch_dars_overall}
	Consider finding penalties for requirements of Figure \ref{fig_ch_dars_ex_vdg}, where $r_4$ is not selected ($x_1=x_2=x_3=1,x_4=0$). From Table~\ref{table_ch_dars_ex_overall} we have $I_{1,4}=I_{3,4}=0.7,I_{2,4}=0.3,I_{4,4}=0.0$. As such, based on~(\ref{eq_ch_dars_penalty}) penalties are computed: $\theta_{1}= \vee(\frac{\lvert 0.0 \rvert +(1-2(1))(0.0)}{2}$, $\frac{\lvert 0.45 \rvert +(1-2(1))(0.5)}{2}$, $\frac{\lvert 0.7 \rvert +(1-2(1))(0.7)}{2}$, $\frac{\lvert 0.7 \rvert +(1-2(0))(0.7)}{2}) =0.7$. Similarly, we have $\theta_2=0.3, \theta_3=0.7$. Therefore, the overall value of the selected requirements $r_1,r_2,r_3$ is derived by~(\ref{eq_ch_dars_ov}) as: $ OV (s_1) = (1-0.7)E(v_1) + (1-0.3)E(v_2) + (1-0.7)E(v_3)$. 
\end{exmp}

%% file: selection_ilp.tex
\subsection{The Integer Programming Model}
\label{ch_dars_selection_ilp}

This section presents our proposed integer programming model for optimizing the overall value of software. The overall value of a software requirement subset, as given by (\ref{eq_ch_dars_ov}), considers user preferences and the impacts of value dependencies on the expected values of the selected requirements. Equations (\ref{Eq_ch_dars_dars})-(\ref{Eq_ch_dars_dars_c5}) give our proposed integer programming model as a main component of DARS. In these equations, $x_i$ is a selection variable denoting whether a requirement $r_i$ is selected ($x_i=1$) or ignored ($x_i=0$). Also $\theta_i$ in (\ref{eq_ch_dars_penalty}) specifies the penalty of a requirement $r_i$, which is the extent to which the expected value of $r_i$ is impacted by ignoring (selecting) requirements with positive (negative) influences on the value of $r_i$. Constraint~(\ref{Eq_ch_dars_dars_c2}) accounts for precedence dependencies among requirements and the value implications of those dependencies. Moreover, for a requirement $r_i$, $\theta_i$ depends on the selection variable $x_j$ and the strength of positive (negative) value dependencies as given by~(\ref{eq_ch_dars_penalty}). As $I_{i,j}$ is computed by (\ref{Eq_ch_dars_influence}), we can restate $\theta_i$ as a function of $x_j$: $\theta_i=f(x_j)$. The objective function~(\ref{Eq_ch_dars_dars}), thus, can be restated as $\text{Maximize } \sum_{i=1}^{n} x_i E(v_i) - x_if(x_j)E(v_i)$ where $x_if(x_j)E(v_i)$ is a quadratic non-linear expression~\cite{boyd2004convex}.

\begin{align}
\label{Eq_ch_dars_dars}
& \text{Maximize } \sum_{i=1}^{n} x_i (1-\theta_i) E(v_i)\\[2pt]
\label{Eq_ch_dars_dars_c1}
&\text{Subject to} \sum_{i=1}^{n} c_i x_i \leq b\\[2pt] 
\label{Eq_ch_dars_dars_c2}
& \begin{cases}
x_i \le x_j  & r_j \text{ precedes } r_i \\[2pt]
x_i \le 1-x_j& r_i \text{ conflicts with } r_j,\text{ }i\neq j= 1,...,n\\[2pt]
\end{cases}\\[2pt]
\label{Eq_ch_dars_dars_c3}
& \theta_{i} \geq \bigg(\frac{\lvert I_{i,j} \rvert + (1-2x_j)I_{i,j}}{2}\bigg), i\neq j = 1,...,n\\[2pt]
\label{Eq_ch_dars_dars_c4}
&\text{ }x_i \in \{0,1\}, i = 1,...,n \\[2pt]
\label{Eq_ch_dars_dars_c5}
&\text{ } 0 \leq \theta_i \leq 1, i = 1,...,n
\end{align}

 Equations (\ref{Eq_ch_dars_dars})-(\ref{Eq_ch_dars_dars_c3}), on the other hand, denote a convex optimization problem as the model maximizes a concave objective function with linear constraints. Convex optimization problems are solvable~\cite{boyd2004convex,mougouei2017integer}. However, for problems of moderate to large sizes, \textit{Integer Linear Programming} (ILP) models are preferred~\cite{luenberger2015linear} as they can be efficiently solved, despite the inherent complexity of NP-hard problems, due to the advances in solving ILP models and availability of efficient tools such as ILOG CPLEX for that purpose. 

\begin{align}
\label{Eq_ch_dars_dars_linear}
&\text{Maximize }  \sum_{i=1}^{n} x_i E(v_i) - y_i E(v_i)\\[2pt]
\label{Eq_ch_dars_dars_linear_c1}
&\text{Subject to} \sum_{i=1}^{n} c_i x_i \leq b\\[2pt]
\label{Eq_ch_dars_dars_linear_c2}
& \begin{cases}
x_i \le x_j  & r_j \text{ precedes } r_i \\[2pt]
x_i \le 1-x_j& r_i \text{ conflicts with } r_j,\text{ }i\neq j= 1,...,n
\end{cases}\\[2pt]
\label{Eq_ch_dars_dars_linear_c3}
& \theta_{i}\geq \bigg(\frac{\lvert I_{i,j} \rvert + (1-2x_j)I_{i,j}}{2}\bigg), i\neq j = 1,...,n\\
\label{Eq_ch_dars_dars_linear_c4}
& -g_i \leq x_i \leq  g_i, i=1,...,n\\[2pt]
\label{Eq_ch_dars_dars_linear_c5}
& 1-(1-g_i) \leq x_i \leq 1+(1-g_i), i=1,...,n\\[2pt]
\label{Eq_ch_dars_dars_linear_c6}
& -g_i \leq y_i \leq g_i, i=1,...,n\\
\label{Eq_ch_dars_dars_linear_c7}
& -(1-g_i)\leq(y_i-\theta_i) \leq (1-g_i), i=1,...,n\\[2pt]
\label{Eq_ch_dars_dars_linear_c8}
&\text{ } 0 \leq y_i \leq 1, i = 1,...,n\\[2pt]
\label{Eq_ch_dars_dars_linear_c9}
&\text{ } 0 \leq \theta_i \leq 1, i = 1,...,n \\[2pt]
\label{Eq_ch_dars_dars_linear_c10}
& \text{ }x_i,g_i \in \{0,1\}, i=1,...,n
\end{align}

This motivates us to consider deriving an ILP version of the model as given by~(\ref{Eq_ch_dars_dars_linear}). In doing so, non-linear expression $x_i\theta_i$ is substituted by linear expression $y_i$ ($y_i=x_i\theta_i$). As such, either $a:(x_i=0,y_i=0)$, or $b:(x_i=1,y_i=\theta_i)$ occur. To capture the relation between $\theta_i$ and $y_i$ in a linear form, we have made use of an auxiliary variable $g_i=\{0,1\}$ and (\ref{Eq_ch_dars_dars_linear_c4})-(\ref{Eq_ch_dars_dars_linear_c8}) are added to the original model. As such, we have either $(g_i=0) \rightarrow a$, or $(g_i=1) \rightarrow b$. Therefore,~(\ref{Eq_ch_dars_dars_linear})-(\ref{Eq_ch_dars_dars_linear_c10}), referred to as DARS-ILP, is linear and can be efficiently solved~\cite{boyd2004convex}, even for large scale requirement sets, by existing commercial solvers such as \textit{IBM CPLEX}. 

%% file: selection_blind.tex
\subsection{The Complementary Model}
\label{ch_dars_selection_blind}

 There might be situations where identifying value dependencies is hard, e.g, when collecting user preferences is costly. In such cases, value dependencies cannot be considered in requirement selection, that may lead to ignoring (selecting) the requirements with significant positive (negative) influences on the values of the selected requirements and therefore result in value loss. When the impacts of negative value dependencies are negligible, value loss induced by ignoring positive value dependencies can be mitigated by trying to select as many requirements as possible, respecting the budget. On this basis, we have proposed a complementary (ILP) model of DARS that reduces the chances that the requirements with positive influences on the values of the selected requirements are ignored. 

\begin{align}
\label{eq_ch_dars_utility_f1}
&f_1(n,x_i,E(v_i)) = \sum_{i=1}^{n} x_i E(v_i)\\
\label{eq_ch_dars_utility_f2}
&f_2 (n,x_i) = \sum_{i=1}^{n} x_i\\
\label{eq_ch_dars_expected}
&E(v_i) = p(r_i)v_i 
\end{align}

Equations (\ref{eq_ch_dars_blind_1})-(\ref{eq_ch_dars_blind_1_c3}) give a multi-objective (bi-objective) formulation of the blind model of DARS, which aims to simultaneously maximize the utility functions $f_1$ and $f_2$ while respecting the budget constraint (\ref{eq_ch_dars_blind_1_c1}) and the precedence constraints (\ref{eq_ch_dars_blind_1_c2}). The utility function $f_1$ in (\ref{eq_ch_dars_utility_f1}) concerns with the expected value of the selected requirements while the utility function $f_2$ specifies the number of the selected requirements as given by (\ref{eq_ch_dars_utility_f2}). In these equations, $b$ denotes the budget and $x_i$ is a decision variable specifying if requirement $r_i$ is selected ($x_i=1$) or not ($x_i=0$). Also, $c_i$ and $E(v_i)$ denote the estimated cost and the expected value of $r_i$ respectively. $E(v_i)$ is computed by (\ref{eq_ch_dars_expected}) where $v_i$ is the estimated value of $r_i$ and $p(r_i)$ denotes the probability that users select $r_i$. The optimization model (\ref{eq_ch_dars_blind_1})-(\ref{eq_ch_dars_blind_1_c3}), aims to find a requirement subset with the highest values for utility functions $f_1$ and $f_2$ while keeping the cost within the budget and respecting the precedence constraints (\ref{eq_ch_dars_blind_1_c2}). However, maximizing the number of the selected requirements (the utility function $f_2$) may conflict with maximizing the expected value of the selected requirements (the utility function $f_1$). 

Hence, finding an optimal subset without knowing the preference of a decision make is not possible; all \textit{Pareto Optimal}~\cite{zhang_multi_objective_2007} subsets found by the optimization model (\ref{eq_ch_dars_blind_1})-(\ref{eq_ch_dars_blind_1_c3}) are considered to be equally good. In a Pareto optimal (\textit{Non-Dominated}) subset found by the model, none of the utility functions $f_1$ or $f_2$ can be improved in value without degrading the other. The optimization model (\ref{eq_ch_dars_blind_1})-(\ref{eq_ch_dars_blind_1_c3}) can be solved in different ways, as discussed in~\cite{marler2004survey}, depending on the viewpoints of the decision makers and, thus, there exist different solution philosophies when solving them. 

\begin{align}
\label{eq_ch_dars_blind_1}
& \text{Maximize } \{f_1(n,x_i,E(v_i)), f_2 (n,x_i)\}\\[5pt]
\label{eq_ch_dars_blind_1_c1}
&\text{Subject to} \sum_{i=1}^{n} x_i c_i \leq b\\[5pt]
\label{eq_ch_dars_blind_1_c2}
& \begin{cases}
x_i \le \text{$x_j$}  & r_j \text{ precedes } r_i \\[5pt]
x_i \le 1-\text{$x_j$}& r_i \text{ conflicts with } r_j,\text{ }i\neq j= 1,...,n\\[5pt]
\end{cases}\\[5pt]
\label{eq_ch_dars_blind_1_c3}
&\text{ }x_i \in \{0,1\}, i = 1,...,n 
\end{align}

We reformulate the complementary model of DARS as a single-objective model in (\ref{eq_ch_dars_blind_2})-(\ref{eq_ch_dars_blind_2_c4}), that aims to avoid ignoring requirements (maximizing $f_2$) as long as the budget constraint (\ref{eq_ch_dars_blind_2_c1}) is respected and the utility function $f_1$ is partly satisfied by guaranteeing a lower-bound $V$ for the expected value of the optimal subset. $V$ will be specified by decision makers. The conflict between the utility functions $f_1$ and $f_2$, thus, is reconciled by maximizing $f_2$ while ensuring a lower-bound for $f_1$. Moreover, precedence dependencies among the requirements are captured by (\ref{eq_ch_dars_blind_2_c3}), where $x_i \le x_j$ states that a requirement $r_i$ requires $r_j$ while $x_i \le(1-x_j)$ means that $r_i$ conflicts with $r_j$.    

\begin{align}
\label{eq_ch_dars_blind_2}
& \text{Maximize } f_1(n,x_i,E(v_i)) \\[5pt]
\label{eq_ch_dars_blind_2_c1}
&\text{Subject to} \sum_{i=1}^{n} x_i c_i \leq b\\[5pt]
\label{eq_ch_dars_blind_2_c2}
& f_2 (n,x_i) \geq V\\[5pt]
\label{eq_ch_dars_blind_2_c3}
& \begin{cases}
x_i \le x_j  & r_j \text{ precedes } r_i \\[5pt]
x_i \le 1-x_j& r_i \text{ conflicts with } r_j,\text{ }i\neq j= 1,...,n\\[5pt]
\end{cases}\\[5pt]
\label{eq_ch_dars_blind_2_c4}
&\text{ }x_i \in \{0,1\}, i = 1,...,n 
\end{align}
\vspace{0.2cm}

Finally, the complementary ILP model of DARS, as given by (\ref{eq_ch_dars_blind_2})-(\ref{eq_ch_dars_blind_2_c4}), is linear and therefore can be efficiently solved~\cite{boyd2004convex}, even for large scale requirement sets, by the existing commercial solvers such as \textit{IBM CPLEX}~\cite{cplex2016v12}. We have implemented, solved, the model using the \textit{Concert Technology} and the \textit{JAVA API of IBM CPLEX}~\cite{cplex2016v12}. 


 
 

%% file: simulation.tex
\section{Simulations}
\label{ch_dars_validation_sim}

This section demonstrates the effectiveness of the ILP model of DARS through simulations. The model accounts for user preferences using the expected values of the requirements ($E(v_i)$) as discussed before. But the interplay between user preferences and value dependencies may interfere with studying the impact of value dependencies on the effectiveness of DARS. This may, particularly, occur when value dependencies are found among frequently preferred requirements with higher values. To merely study the impacts of value dependencies, we have substituted, without loss of generality, the expected values of requirements $(E(v_i))$ with their corresponding estimated values ($v_i$) in (\ref{Eq_ch_dars_dars_linear})-(\ref{Eq_ch_dars_dars_linear_c10}), thus factoring out the impact of user preferences from simulations. 

We have further, considered different budgets and levels of precedence dependencies in simulations. Simulations were carried out for $27$ requirements of a real-world software with the estimated values and costs scaled into $[0,20]$ (Table~\ref{table_ch_dars_cost_value}). Value dependencies and precedence dependencies among the requirements were randomly generated. The Java API of IBM CPLEX was used to implement the optimization models of the BK, PCBK, and DARS. Also, the callable library ILOG CPLEX 12.6.2 was used to run these models on a machine with a Core i7-2600 3.4 GHz processor and 16 GB of RAM. 

\begin{table}[!htbp]
	\caption{The estimated costs and values of the requirements.}
	\label{table_ch_dars_cost_value}
	\centering
	\input{table_cost_value}
\end{table}

Requirement selection was performed for different percentages of budget ($\%\text{Budget}=\{1,...,100\}$), value dependency levels (VDL $\in [0,1]$), negative value dependency levels (NVDL $\in [0,1]$), precedence dependency levels (PDL $\in [0,1]$), and negative precedence dependency levels (NPDL $\in [0,1]$). PDL and NPDL were computed by (\ref{Eq_pdl}) and (\ref{Eq_npdl}) respectively to build random precedence graphs $G$ with $n=27$ nodes (requirements). In these equations, $k$ and $j$ denote the total number of precedence dependencies and the number of negative precedence dependencies respectively.

\vspace{-0.2cm}
\begin{align}
\label{Eq_pdl}
&PDL(G)=\frac{k}{\Perm{n}{2}}=\frac{k}{n (n-1)} \\
\label{Eq_npdl}
&NPDL(G)=\frac{j}{k}
\end{align}

Table \ref{table_sim_design} lists our simulation settings designed to answers the following research questions. 

\input{rq_dars_ilp_sim}

To simulate value dependencies for a desired VDL and NVDL, uniformly distributed random numbers in $[-1,1]$ were generated, where the sign and magnitude of each number specified the quality and the strength of its corresponding explicit value dependency respectively. In a similar way, for a desired PDL and NPDL, random numbers in $\{-1,0,1\}$ were generated where $1$ ($-1$) specified a positive (negative) precedence dependency and $0$ denoted the absence of any precedence dependency from a requirement $r_i$ to $r_j$. Furthermore, percentages of the overall value of selected requirements ($\%$OV$=\frac{\text{OV}}{\sum_{i=1}^{27}v_i}$) were used to measure the performance of the simulated selection methods. 

The performance of BK was, however, arbitrary as it does not consider precedence dependencies and therefore in many cases, depending on the PDL and NPDL, violates those dependencies giving infeasible solutions with no value ($\%\text{OV}=0$). On the other hand, the PCBK method enhances BK by considering precedence dependencies. As such, the PCBK method always outperforms the BK method giving higher or equal $\%$OV. Hence, we have mainly focused on comparing the performance of the PCBK and DARS methods. Moreover, Increase-Decrease methods were not simulated as they do not specify how to achieve the amount of increased or decreased values (Section~\ref{sec_related}). 


\begin{table}[!htbp]
	\caption{Performance simulations for the selection methods.}
	\label{table_sim_design}
	\centering
	\input{table_sim_design_all}
\end{table}



%% file: table_cost_value.tex
\scriptsize\resizebox {0.4\textwidth }{!}{
\begin{tabular}{lllllll}
	\toprule[1.5pt]
	\cellcolor[HTML]{FFFFFF}{\color[HTML]{000000} \textbf{$r_i$}} & \cellcolor[HTML]{FFFFFF}{\color[HTML]{000000} \textbf{$c_i$}} & \cellcolor[HTML]{FFFFFF}{\color[HTML]{000000} \textbf{$v_i$}} & \cellcolor[HTML]{FFFFFF}{\color[HTML]{000000} \textbf{$r_i$}} & \cellcolor[HTML]{FFFFFF}{\color[HTML]{000000} \textbf{$c_i$}} & \cellcolor[HTML]{FFFFFF}{\color[HTML]{000000} \textbf{$v_i$}} \\ \midrule
	$r_1$ & 05.00 & 10.00 & $r_{15}$ & 15.00 & 08.00 \\
	$r_2$ & 20.00 & 20.00 & $r_{16}$ & 13.00 & 10.00 \\
	$r_3$ & 00.00 & 04.00 & $r_{17}$ & 14.00 & 06.00  \\
	$r_4$ & 10.00 & 17.00 & $r_{18}$ & 03.00 & 10.00  \\
	$r_5$ & 01.00 & 03.00 & $r_{19}$ & 10.00 & 20.00  \\
	$r_6$ & 20.00 & 20.00 & $r_{20}$ & 07.00 & 20.00  \\
	$r_7$ & 06.00 & 15.00 & $r_{21}$ & 12.00 & 15.00  \\
	$r_8$ & 05.00 & 09.00 & $r_{22}$& 15.00 & 20.00  \\
	$r_9$ & 16.00 & 20.00 & $r_{23}$ & 08.00  & 20.00  \\
	$r_{10}$ & 10.00 & 16.00 & $r_{24}$ & 02.00  & 05.00 \\
	$r_{11}$ & 04.00 & 20.00 & $r_{25}$ & 10.00   & 00.00 \\
	$r_{12}$ & 03.00 & 10.00 &$r_{26}$ & 00.00 & 00.00  \\
	$r_{13}$ & 05.00 & 06.00 &$r_{27}$ & 01.00 & 00.00 \\
	$r_{14}$ & 07.00 & 08.00 &        &  &  \\[2pt] \midrule
	{\color[HTML]{000000} \textbf{Sum}} & {\color[HTML]{000000} 112.00} & {\color[HTML]{000000} 178.00} & {\color[HTML]{000000} -} & {\color[HTML]{000000} 110.00} & {\color[HTML]{000000} 134.00} \\ \bottomrule[1.5pt]
\end{tabular}

}

%% file: rq_dars_ilp_sim.tex
\begin{itemize}[leftmargin=1.8cm]
	\itemsep0em 
	\item[(\textbf{RQ1})] What is the impact of value dependencies on the performance of DARS in the presence of various budget constrains?  
	\item[(\textbf{RQ2})] What is the impact of negative value dependencies on the performance of DARS in the presence of various budget constrains?  
	\item[(\textbf{RQ3})] What is the impact of precedence dependencies on the performance of DARS in the presence of various budget constrains? 
	\item[(\textbf{RQ4})] What is the impact of negative precedence dependencies on the performance of DARS in the presence of various budget constrains?
	\item[(\textbf{RQ5})] What is the impact of negative value dependencies on the performance of DARS in the presence of different levels of value dependencies?
	\item[(\textbf{RQ6})] What is the impact of negative precedence dependencies on the performance of DARS for different levels of precedence dependencies?
\end{itemize}

%% file: table_sim_design_all.tex
\resizebox {0.4\textwidth }{!}{
	\begin{tabular}{clllll}
	\toprule[1.5pt]
	\rowcolor{gray!30}
	\textbf{\cellcolor{black}\textcolor{white}{Simulation}} &
	\textbf{\cellcolor{black}\textcolor{white}{$\%$Budget}} &
	\textbf{\cellcolor{black}\textcolor{white}{VDL}} &
	\textbf{\cellcolor{black}\textcolor{white}{NVDL}} &
    \textbf{\cellcolor{black}\textcolor{white}{PDL}} &
	\textbf{\cellcolor{black}\textcolor{white}{NPDL}} 
	\bigstrut\\
	\rowcolor{gray!25}
	I &
	[0,100] &
	[0,1] &
	0.00&
	0.02 &
	0.00
	\bigstrut\\
	II &
	[0,100] &
	0.15 &
	[0,1] &
	0.02 &
	0.00
	\bigstrut\\
	\rowcolor{gray!25}
	III &
	[0,100] &
	0.15 &
	0.00&
	[0,1] &
	0.00
	\bigstrut\\
	IV &
	[0,100] &
	0.15 &
	0.00&
	0.02 &
	[0,1]
	\bigstrut\\
	\rowcolor{gray!25}
	V &
	95 &
	[0,1] &
	[0,1] &
	0.02 &
	0.00
	\bigstrut\\
	VI &
	95 &
	0.15 &
	0.00&
	[0,1] &
	[0,1]
	\bigstrut\\
	\bottomrule[1.5pt]
\end{tabular}
}

%% file: simulation_I.tex
\subsection{Value Dependencies vs Budget}

\begin{figure}[htbp]
	\begin{center}
		\subfigure[$\%$AV(BK)]{%
			\label{fig_ch_dars_sim_0_av_0}
			\includegraphics[scale=0.278]{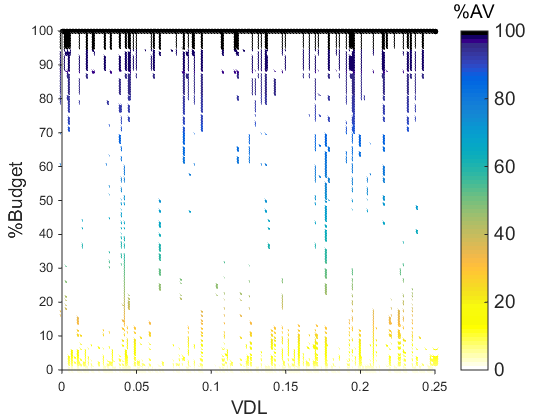}
		}
		\subfigure[$\%$OV(BK)]{%
			\label{fig_ch_dars_sim_0_ov_0}
			\includegraphics[scale=0.278]{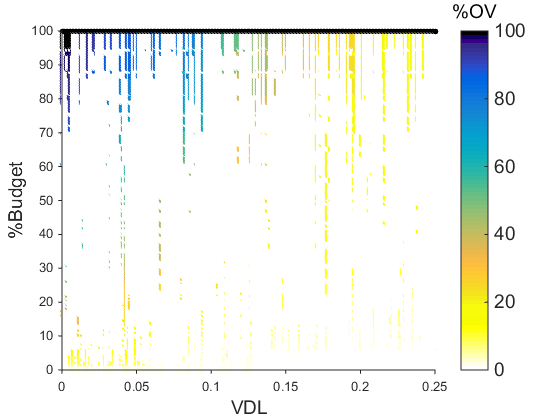}
		}\\
		\subfigure[$\%$AV(PCBK)]{%
			\label{fig_ch_dars_sim_0_av_1}
			\includegraphics[scale=0.278]{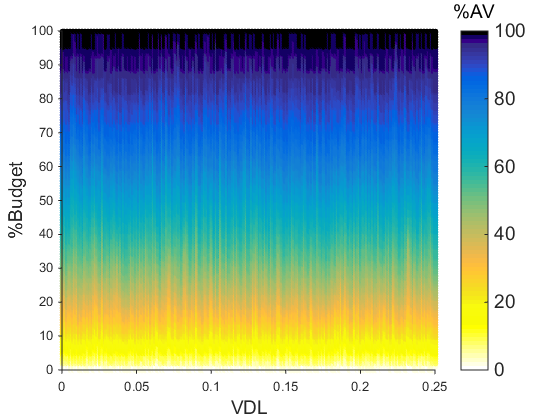}
		}
		\subfigure[$\%$OV(PCBK)]{%
			\label{fig_ch_dars_sim_0_ov_1}
			\includegraphics[scale=0.278]{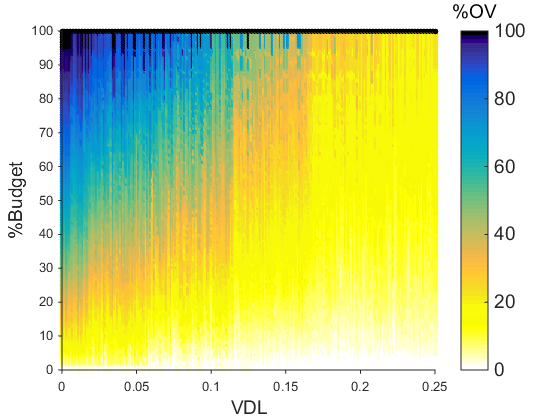}
		}\\
		\subfigure[$\%$AV(DARS)]{%
			\label{fig_ch_dars_sim_0_av_2}
			\includegraphics[scale=0.278]{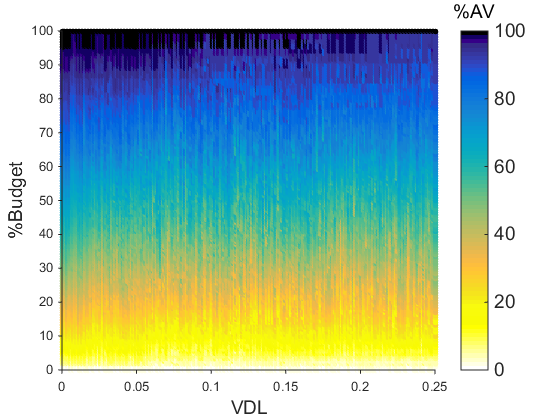}
		}
		\subfigure[$\%$OV(DARS)]{%
			\label{fig_ch_dars_sim_0_ov_2}
			\includegraphics[scale=0.278]{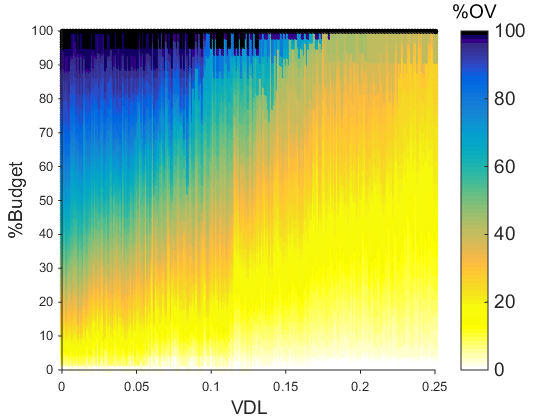}
		}
	\end{center}
	\caption{%
		Simulation I ($\%$Budget vs. VDL).
	}%
	\label{fig_ch_dars_sim_0}
\end{figure}

To answer (\textbf{RQ1}), Simulation I in Table~\ref{table_sim_design} was carried out for $\%$Budget $\in [0,100]$ and $VDL = [0,1]$. Figure~\ref{fig_ch_dars_sim_0} shows the percentages of accumulated value ($\%$AV) and overall value ($\%$OV) achieved from the selection methods. As expected, BK ignored/violated precedence dependencies and generated infeasible solutions with $\%$AV=$\%$OV$=0$ in most cases (Figure~\ref{fig_ch_dars_sim_0_av_0} and Figure~\ref{fig_ch_dars_sim_0_ov_0}). It is clear that with no precedence dependencies (PDL=0), BK and PCBK will perform the same. We further, observed (Figure~\ref{fig_ch_dars_sim_0}) that for a given $\%$Budget and NVLD=0 increasing VDL generally decreased $\%$OV achieved by all selection methods. The reason is increasing VDL increases the chances that the positive dependencies of a requirement are ignored, that may negatively influence the values of the selected requirements.

\begin{figure}[!htb]
	\begin{center}
		\subfigure[$\%\Delta$OV(DARS,BK)]{%
			\label{fig_ch_dars_sim_diff_0_ov_20}
			\includegraphics[scale=0.278]{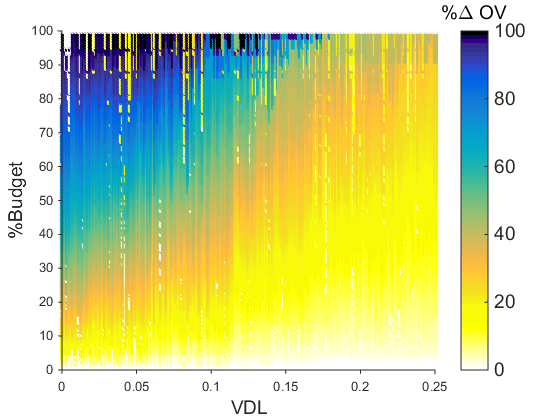}
		}
		\subfigure[$\%\Delta$OV(DARS,PCBK)]{%
			\label{fig_ch_dars_sim_diff_0_ov_21}
			\includegraphics[scale=0.278]{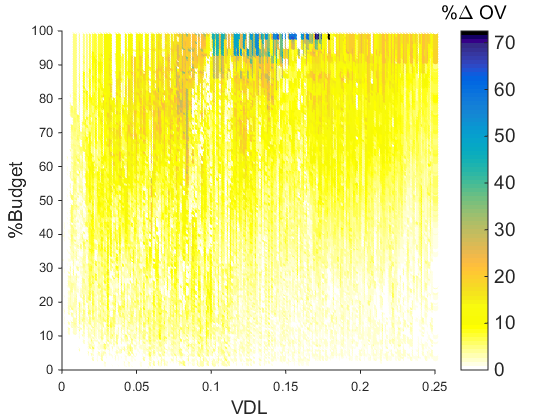}
		}
		\subfigure[$\%\Delta$AV(DARS,BK)]{%
			\label{fig_ch_dars_sim_diff_0_av_20}
			\includegraphics[scale=0.278]{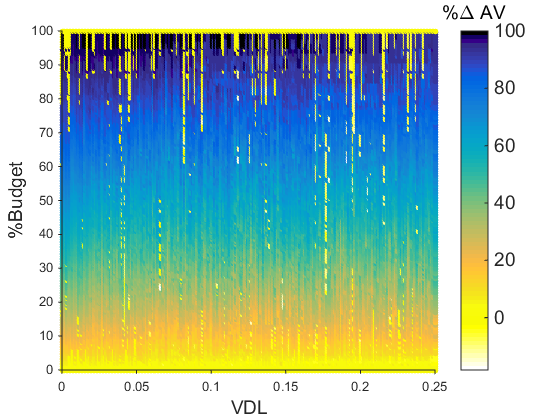}
		}
		\subfigure[$\%\Delta$AV(DARS,PCBK)]{%
			\label{fig_ch_dars_sim_diff_0_av_21}
			\includegraphics[scale=0.278]{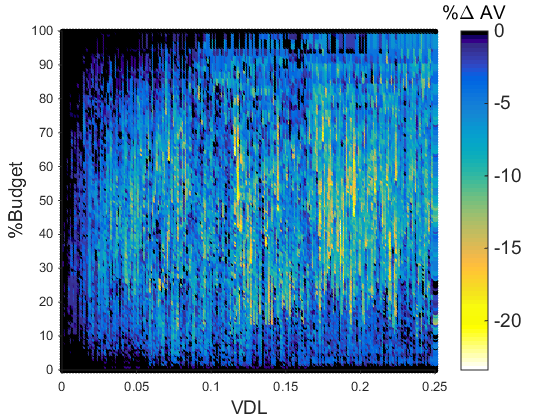}
		}
	\end{center}
	\caption{%
		Simulation I ($\%$Budget vs. VDL).
	}%
	\label{fig_ch_dars_sim_diff_0}
\end{figure}

Figure~\ref{fig_ch_dars_sim_0} and Figure~\ref{fig_ch_dars_sim_diff_0}, also, show that DARS gives higher $\%$OV for all VDLs and $\%$Budget compared to the PCBK and BK methods. The reason is the ILP model of DARS considers value dependencies as well as the value implications of precedence dependencies while the BK method ignores dependencies all together, and the PCBK method only considers precedence dependencies. Figure~\ref{fig_ch_dars_sim_diff_0} compares $\%$OV and $\%$AV provided by DARS against those of BK and PCBK for various $\%$Budget and VDLs. 
We have $\%\Delta$OV($m_1$,$m_2$)=$\%$OV($m_1$)$-\%$OV($m_2$) and $\%\Delta$AV($m_1$,$m_2$)=$\%$AV($m_1$)$-\%$AV ($m_2$) for selection methods $m_1$ and $m_2$. Our results demonstrated that DARS outperformed BK and PCBK by providing higher $\%$OV (Figure~\ref{fig_ch_dars_sim_diff_0}). Moreover, we observed that finding a requirement subset with the highest accumulated value conflicts with finding a subset with the highest overall value: to maximize OV and AV are conflicting objectives. This is demonstrated in many points in the graphs of Figure~\ref{fig_ch_dars_sim_diff_0_av_21} and Figure \ref{fig_ch_dars_sim_diff_0_ov_21}, where for a given $\%$Budget and VDL, $\%\Delta$AV(DARS,PCBK) $<0$ while $\%\Delta$OV(DARS,PCBK) $>0$. 

%% file: simulation_II.tex
\subsection{Negative Value Dependencies vs Budget}
\label{sim_ii}
To answer (\textbf{RQ2}), Simulation II was performed for various budgets and NVDLs with settings in Table~\ref{table_sim_design}. We observed that increasing NVDL resulted in arbitrary changes in $\%$OV achieved from PCBK and DARS. This is specified by (\ref{Eq_ch_dars_influence}) where increasing NVDL may arbitrarily increase or decrease the penalty of selecting or ignoring requirements depending on the strengths of positive and negative dependencies and the structure of the value dependency graph of the requirements.

\begin{figure}[!htb]
    \begin{center}
        \subfigure[$\%$OV(PCBK)]{%
            \label{fig_ch_dars_sim_1_ov_1}
            \includegraphics[scale=0.278]{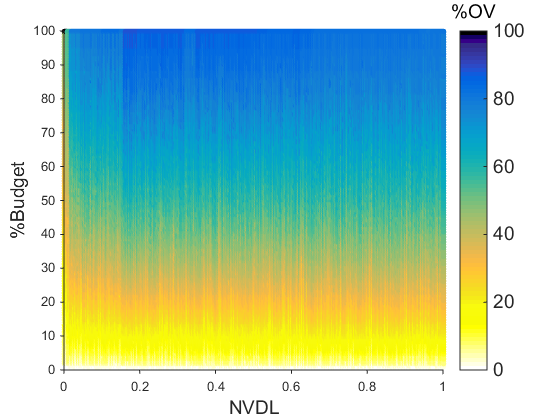}
        }
        \subfigure[$\%$OV(DARS)]{%
            \label{fig_ch_dars_sim_1_ov_2}
            \includegraphics[scale=0.278]{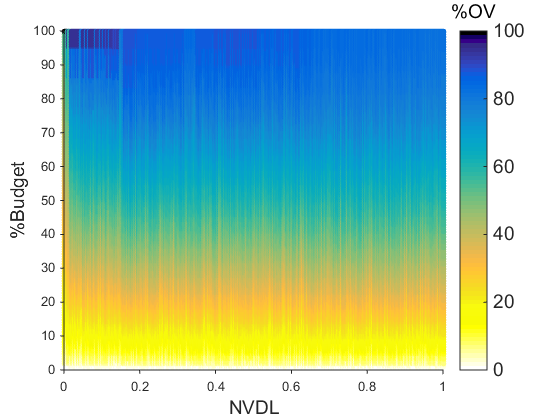}
        }
        \subfigure[$\%\Delta$OV(DARS,PCBK)]{%
            \label{fig_ch_dars_sim_diff_1_ov_21}
            \includegraphics[scale=0.278]{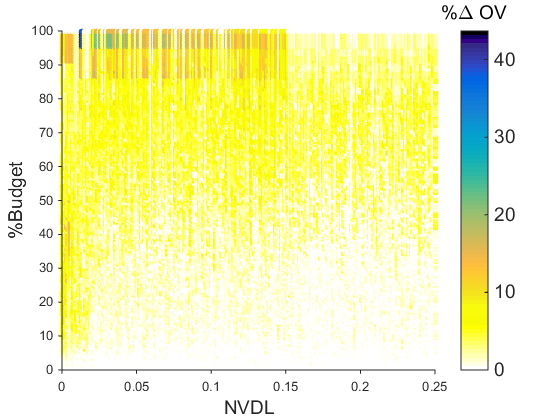}
        }
    \end{center}
    \caption{%
            Simulation II ($\%$Budget vs. NVDL).
    }%
    \label{fig_ch_dars_sim_1}
\end{figure}

Moreover, for $\%$Budget=100 and $NVDL \rightarrow 0$, we observed (Figure \ref{fig_ch_dars_sim_1}) that the maximum $\%$OV can be achieved in both PCBK and DARS methods. The reason is in such cases positive value dependencies do not matter as no requirement is excluded from the optimal subset due to the presence of sufficient budget.

\vspace{0,25cm}

Nonetheless, with NVDL increase in Figure \ref{fig_ch_dars_sim_1} the maximum $\%$OV cannot be achieved even with sufficient budget. The reason is in such cases even selecting requirements may reduce the values of other requirements due to negative value dependencies among them. We further, observed (Figure \ref{fig_ch_dars_sim_diff_1_ov_21}) that, consistent with Simulation I, DARS always provided higher $\%$OV compared to PCBK. Also, the BK method failed to find feasible solutions in most simulations due to violating precedence dependencies.  



%% file: simulation_III.tex
\subsection{Precedence Dependencies vs Budget}
	
\begin{figure}[!htb]
	\begin{center}
		\subfigure[$\%$OV(PCBK)]{%
			\label{fig_ch_dars_sim_2_ov_1}
			\includegraphics[scale=0.278]{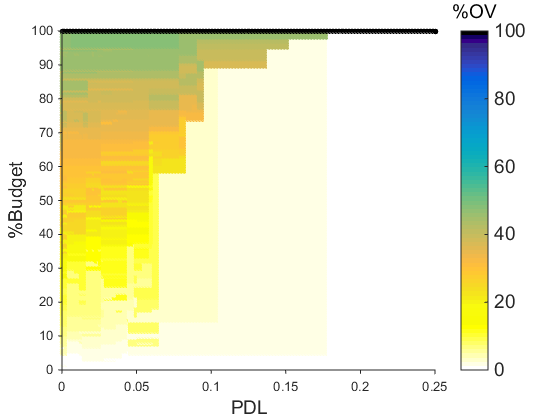}
		}
		\subfigure[$\%$OV(DARS)]{%
			\label{fig_ch_dars_sim_2_ov_2}
			\includegraphics[scale=0.278]{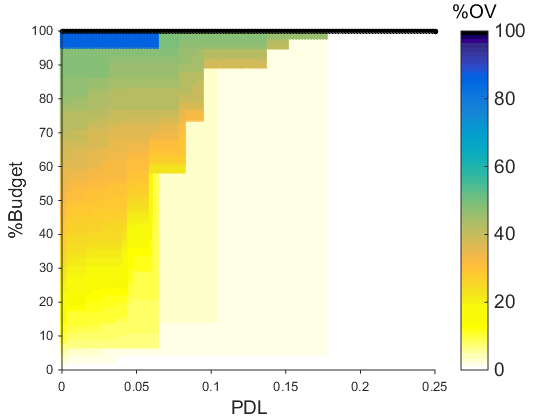}
		}
	    \subfigure[$\%\Delta$OV(DARS,PCBK)]{%
	    	\label{fig_ch_dars_sim_diff_2_ov_21}
	    	\includegraphics[scale=0.278]{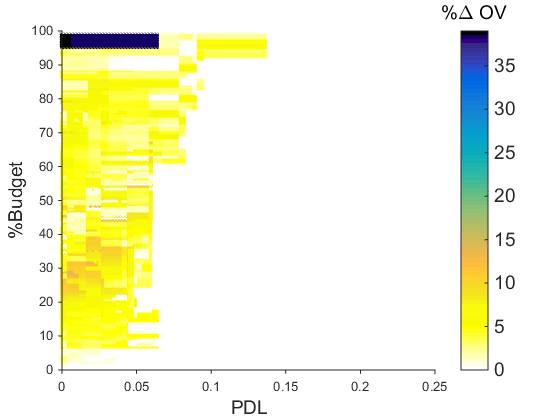}
	    }
	\end{center}
	\caption{%
			Simulation III ($\%$Budget vs. PDL).
	}%
	\label{fig_ch_dars_sim_2}
\end{figure}

To answer (\textbf{RQ3}), Simulation III was performed for various $\%$Budget and PDLs with settings in Table~\ref{table_sim_design}. For a given VDL, we observed that increasing PDL generally resulted in decreasing the $\%$OV achieved from DARS (Figure~\ref{fig_ch_dars_sim_2_ov_2}). The reason is increasing PDL reduces the number of feasible solutions that maintain the precedence constraints. This is also described as the selection deficiency problem (SDP) \cite{mougouei2016factoring} where the efficiency of selection models is constrained by precedence dependencies. A similar effect was observed for PCBK (Figure~\ref{fig_ch_dars_sim_2_ov_1}). 

However, decreasing $\%$OV with PDL increase did not monotonically occur when PCBK was used; increasing PDL resulted in higher $\%$OV in some cases. Such arbitrary effects are more tangible in simulations with $\%\text{Budget} \geq 70$ and $vdl \leq 0.1$. The reason is PCBK ignores value dependencies. As such, even with fewer precedence constraints, PCBK may choose a solution with lower $\%$OV. On the contrary, DARS gave higher $\%$OV for smaller PDLs (Figure~\ref{fig_ch_dars_sim_2_ov_2}). Figure~\ref{fig_ch_dars_sim_diff_2_ov_21} shows DARS outperformed PCBK. But, for PDL $\geq 0.15$, the number of feasible solutions was significantly reduced in both methods and the their performances converged, DARS lost its advantage.

%% file: simulation_IV.tex
\begin{figure}[h]
	\begin{center}
		\subfigure[$\%$OV(PCBK)]{%
			\label{fig_ch_dars_sim_3_ov_1}
			\includegraphics[scale=0.278]{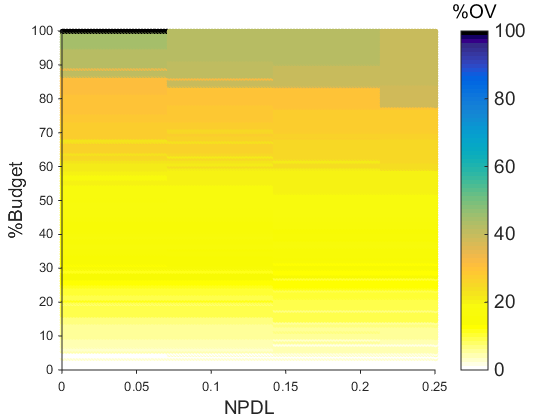}
		}
		\subfigure[$\%$OV(DARS)]{%
			\label{fig_ch_dars_sim_3_ov_2}
			\includegraphics[scale=0.278]{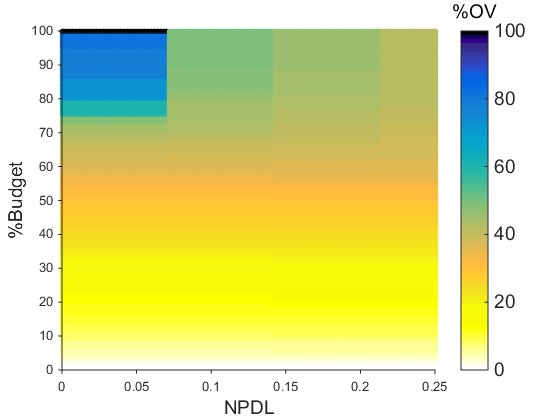}
		}
		\subfigure[$\%\Delta$OV(DARS,PCBK)]{%
			\label{fig_ch_dars_sim_diff_3_ov_21}
			\includegraphics[scale=0.278]{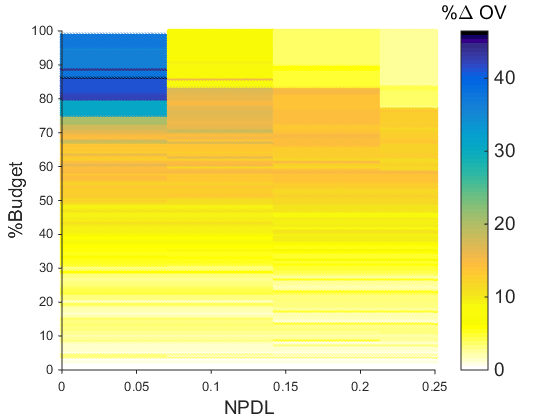}
		}
	\end{center}
	\caption{%
			Simulation IV ($\%$Budget vs. NPDL).
	}%
	\label{fig_ch_dars_sim_3}
\end{figure}

\subsection{Negative Precedence Dependencies vs Budget}
\label{sim_iv}

To answer (\textbf{RQ4}), Simulation IV was performed for different $\%$Budget and NPDLs. For both PCBK and DARS, increasing NPDL limited the number of feasible solutions while increasing the chances that certain requirements were selected. The former resulted in decreasing $\%$OV while the latter arbitrarily increased $\%$OV (\ref{fig_ch_dars_sim_3}); consider requirements $r_1$, $r_2$, $r_3$, where there are negative precede dependencies from $r_1$ to $r_2$ ($x_1 \leq (1-x_2)$) and from $r_2$ to $r_3$, ($x_2 \leq (1-x_3)$): selecting $r_3$ results in ignoring $r_2$, which increases the chances that $r_1$ is selected. 

We further, observed (Figure~\ref{fig_ch_dars_sim_diff_3_ov_21}) that DARS outperformed PCBK for up to around 46$\%$. Moreover, for $75 \leq \%\text{Budget} < 100$ and $NPDL \leq 0.15$, we observed that $\Delta(\text{PCBK},\text{BK})$ and $\Delta(\text{DARS},\text{PCBK})$ were the highest. The reason is DARS finds better solutions in the presence of fewer negative precedence constraints and more budget.





%% file: simulation_V.tex
\vspace{0.5cm}
\begin{figure}[h]
	\begin{center}
		\subfigure[$\%$OV(PCBK)]{%
			\label{fig_ch_dars_sim_4_ov_1}
			\includegraphics[scale=0.278]{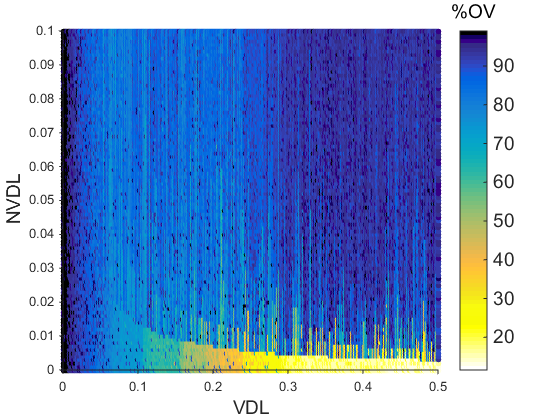}
		}
		\subfigure[$\%$OV(DARS)]{%
			\label{fig_ch_dars_sim_4_ov_2}
			\includegraphics[scale=0.278]{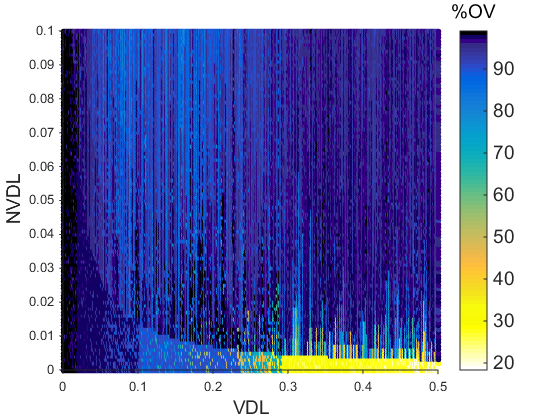}
		}
		\subfigure[$\%\Delta$OV(DARS,PCBK)]{%
			\label{fig_ch_dars_sim_diff_4_ov_21}
			\includegraphics[scale=0.278]{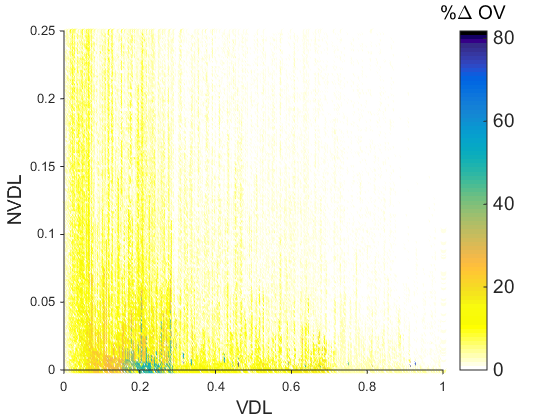}
		}
	\end{center}
	\caption{%
		Simulation V (NVDL vs. VDL).
	}%
	\label{fig_ch_dars_sim_4}
\end{figure}

\vspace{-0.278cm}
\subsection{Positive vs Negative Value Dependencies}
To answer (\textbf{RQ5}), Simulation V was performed for different VDLs and NVDL with $\%\text{Budget}=95$, PDL=$0.02$, and NPDL =$0.0$ as given in Table~\ref{table_sim_design}. The impact of VDL is shown to vary for smaller and larger NVDLs. For smaller NVDLs ($NVDL \leq 0.01$) increasing VDL monotonically decreased the $\%$OV provided by the selection methods. In other cases, however, increasing VDL was demonstrated to increase $\%$OV, although such increase was not monotonic. The reason is, as explained earlier, higher NVDLs increase the chances that simultaneous negative and positive value dependencies from a requirement $r_i$ to $r_j$ exist and therefore, negative value dependencies from $r_i$ to $r_j$ ($\rho(r_i,r_j)^{-\infty}$) mitigate the impact of positive dependencies from $r_i$ to $r_j$ ($\rho(r_i,r_j)^{+\infty}$) and vice versa. This reduces the overall influence of $r_j$ on $r_i$ based on (\ref{Eq_ch_dars_influence}). Hence, neither selecting nor ignoring $r_j$ does result in a significant loss in the value of $r_i$ and the $\%$OV of the selected subset of the requirements.

%% file: simulation_VI.tex
\subsection{Positive vs Negative Precedence Dependencies}
\label{sim_vi}

\begin{figure}[!htb]
	\begin{center}
		\subfigure[$\%$OV(PCBK)]{%
			\label{fig_ch_dars_sim_9_ov_1}
			\includegraphics[scale=0.278]{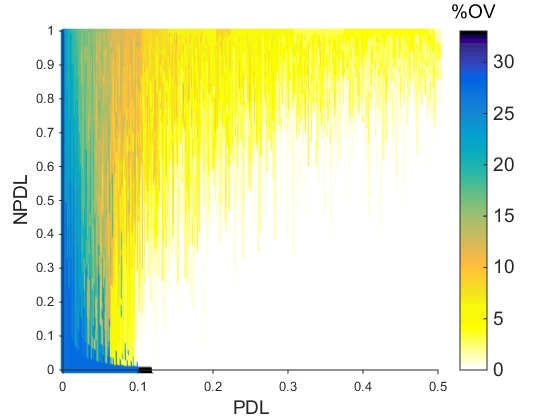}
		}
		\subfigure[$\%$OV(DARS)]{%
			\label{fig_ch_dars_sim_9_ov_2}
			\includegraphics[scale=0.278]{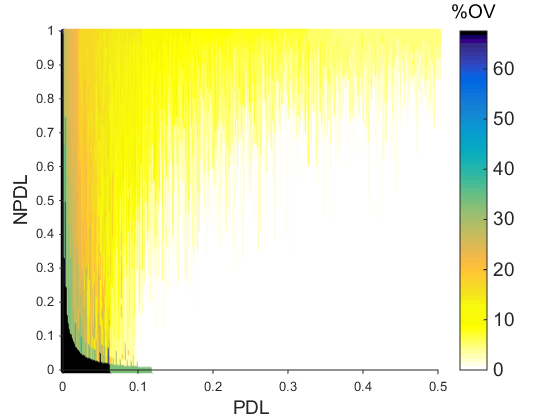}
		}
	    \subfigure[$\%\Delta$OV(DARS,PCBK)]{%
	    	\label{fig_ch_dars_sim_diff_9_ov_21}
	    	\includegraphics[scale=0.278]{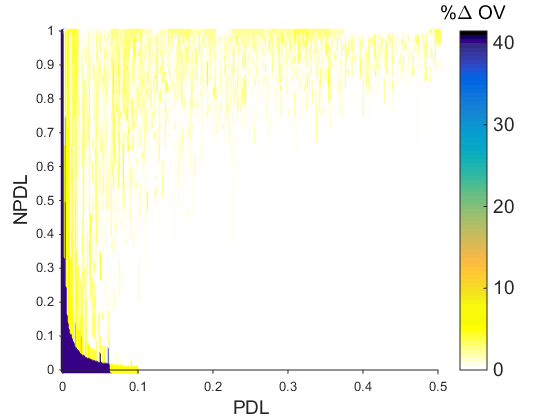}
	    }
	\end{center}
	\caption{%
		Simulation VI (NPDL vs. PDL).
	}%
	\label{fig_ch_dars_sim_9}
\end{figure}

To answer (\textbf{RQ6}), simulations for different PDLs and NPDLs were performed with settings of Simulation VI in Table~\ref{table_sim_design}. Our simulations showed (Figure~\ref{fig_ch_dars_sim_9}) that increasing PDL, in general, decreased the $\%$OV provided by the PCBK and DARS methods. This decrease, nevertheless, was not monotonic in the presence of negative precedence dependencies (NPDL$\neq 0$). The reason for this arbitrary impact of negative precedence dependencies was explained in detail in Section~\ref{sim_iv}. Our simulations showed that for any PDL, there exists a threshold $t$, where PCBK and DARS do not give any value for NPDLs $<t$. These thresholds increased as the PDL increased. This is more visible for PDL $\geq 0.1$. To further explain this, consider a requirement set $R=\{r_1,r_2\}$ with equal costs and equal values, where $r_1$ requires $r_2$ (positive precedence dependency from $r_1$ to $r_2$) and $r_2$ requires $r_1$. This means we have $PDL=1$ and $NPDL=0$. As such, for $\%\text{Budget}<100$, either or both of the $r_1$ and $r_2$ will have to be excluded from the optimal subset which results in violating precedence dependencies and therefore no feasible solution can be found resulting in $\%$OV=0. 

However, for $NPDl=0.5$, one of the precedence dependencies will change to negative (conflicts-with). As this is performed randomly, we have either (a): $r_1$ requires $r_2$ AND $r_2$ conflicts with $r_1$ or (b): $r_2$ requires $r_1$ AND $r_1$ conflicts with $r_2$. It is clear that in either case (for Budget $\geq 50$), at least one requirement ($r_2$ in (a) and $r_1$ in (b)) can be selected. For $\%\text{Budget}=100$, nevertheless, both $r_1$ and $r_2$ are selected in (a). This, clearly shows how, for a given $PDL$, increasing $NPDL$ can provide higher $\%OV$. 

Last but not least, we observed that for simulations with PDL$\geq 0.1$, the performance of PCBK and DARS converged as both methods provided similar $\%OV$: $\Delta\%$OV(DARS,PCBK)$\rightarrow 0$. But this was not the case in the presence of higher levels of negative precedence dependencies. The reason is a large number of precedence dependencies substantially reduces the number of feasible solutions impacting the performance of DARS and PCBK. Nonetheless, increasing NPDL can increase the number of feasible solutions as explained above. Under such circumstances, it is clear that the DARS method can make better choices with regard to the $\%$OV as it takes into account value dependencies in addition to the precedence dependencies.  

%% file: scalability.tex
\section{Scalability Analysis}
\label{ch_dars_scalability}
This section evaluates the scalability of DARS for integrating value dependencies in software requirement selection. We generate random datasets with different numbers of requirements to investigate the scalability of the ILP model of DARS (DARS-ILP) for different scenarios in relation to value dependencies as well as precedence dependencies among requirements. Simulations thus were designed to answer the following questions. 

\input{rq_dars_ilp_scalability}
 
%
%

The ILP model of DARS, as given by~(\ref{Eq_ch_dars_dars_linear})-(\ref{Eq_ch_dars_dars_linear_c8}), is scalable to datasets with a large number of requirements, different budget constraints, and various degrees of precedence/value dependencies. To demonstrate this, runtime simulations in Table~\ref{table_sim_scalability_design} were carried out. To simulate value dependencies for a desired VDL and NVDL, uniformly distributed random numbers in $[-1,1]$ were generated, where the sign and magnitude of each number specified the quality and the strength of its corresponding explicit value dependency. We used PDL and NPDL as given by (\ref{Eq_pdl}) and (\ref{Eq_npdl}) to specify the degree of precedence dependencies in a precedence graph of the requirements. 

\begin{table}[!htbp]
	\caption{Runtime simulations}
	\label{table_sim_scalability_design}
	\centering
	\input{table_sim_scalability_design}
\end{table}


For a given PDL and NPDL, random numbers in $\{-1,0,1\}$ were generated where $1$ ($-1$) specified a positive (negative) precedence dependency and $0$ denoted the absence of any precedence dependency from a requirement $r_i$ to $r_j$. Simulations were carried out using the callable library ILOG CPLEX 12.6.2 on a windows machine with a Core i7-2600 3.4 GHz processor and 16 GB of RAM.

\begin{figure}[htbp]
	\begin{center}
		\subfigure[Simulation 1]{%
			\label{fig_ch_dars_t_n_log}
			\includegraphics[scale=0.2785]{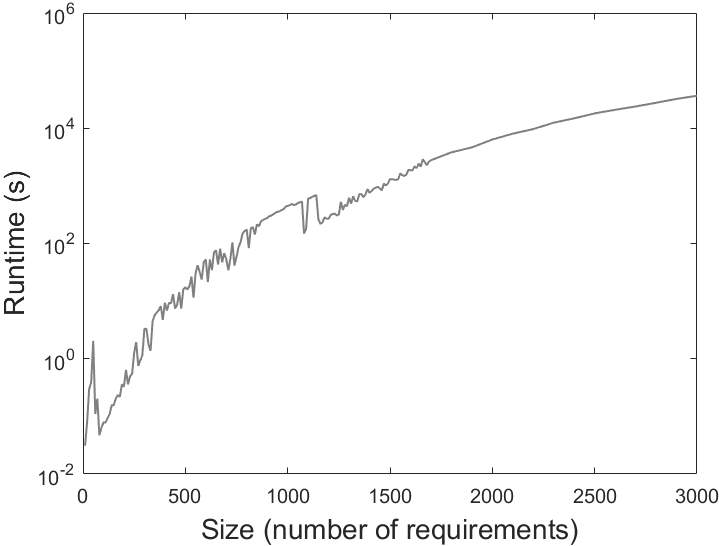}
		}
		\subfigure[Simulation 2]{%
			\label{fig_ch_dars_t_b}
			\includegraphics[scale=0.2785]{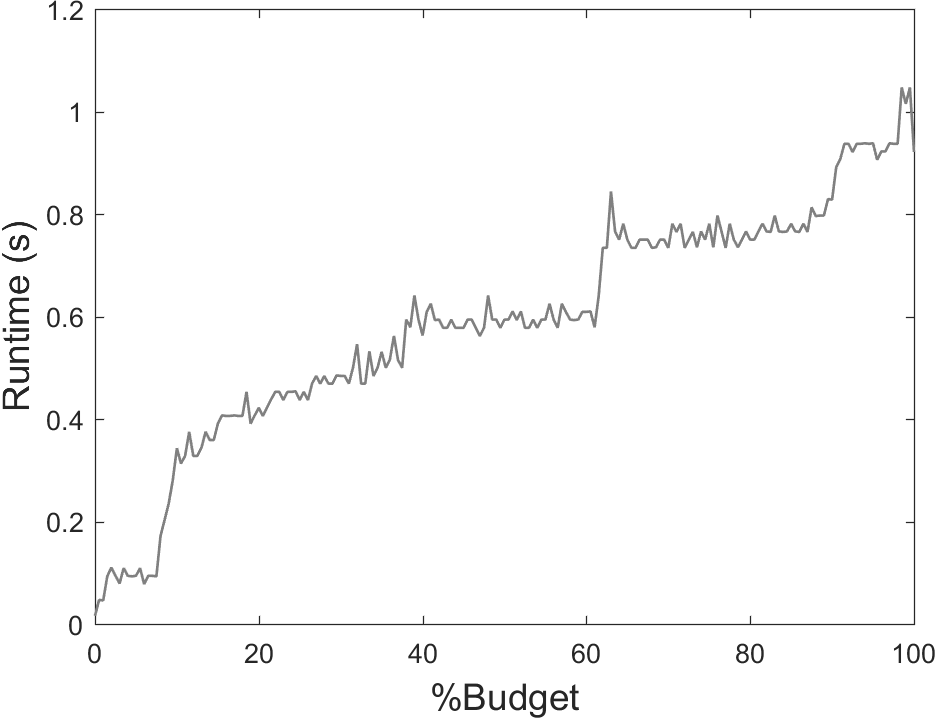}
		}\\
		\subfigure[Simulation 3]{%
			\label{fig_ch_dars_t_pdl}
			\includegraphics[scale=0.2785]{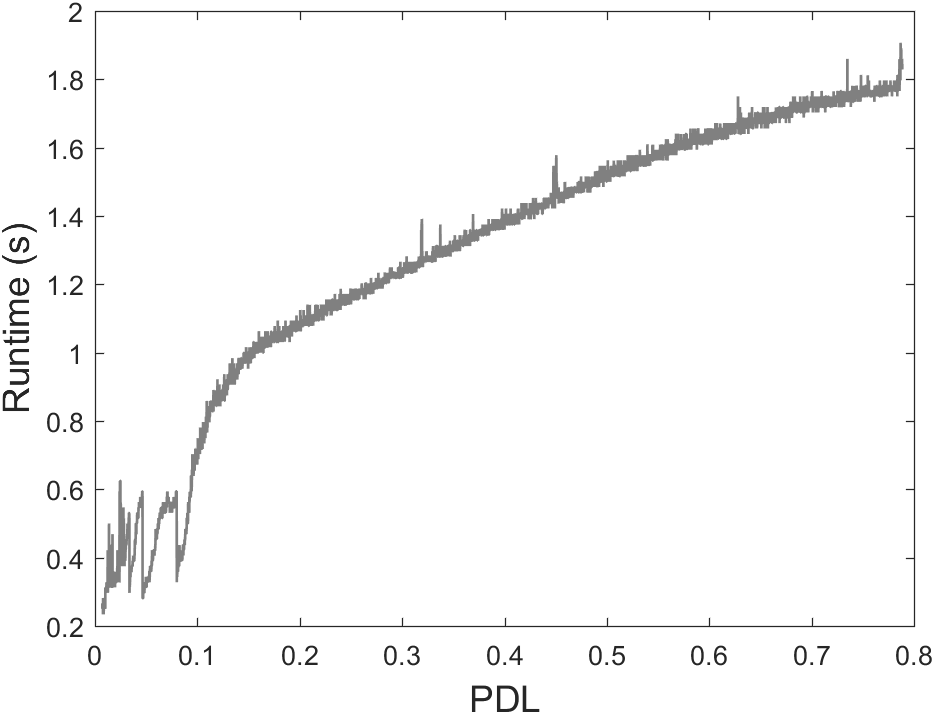}
		}
		\subfigure[Simulation 4]{%
			\includegraphics[scale=0.2785]{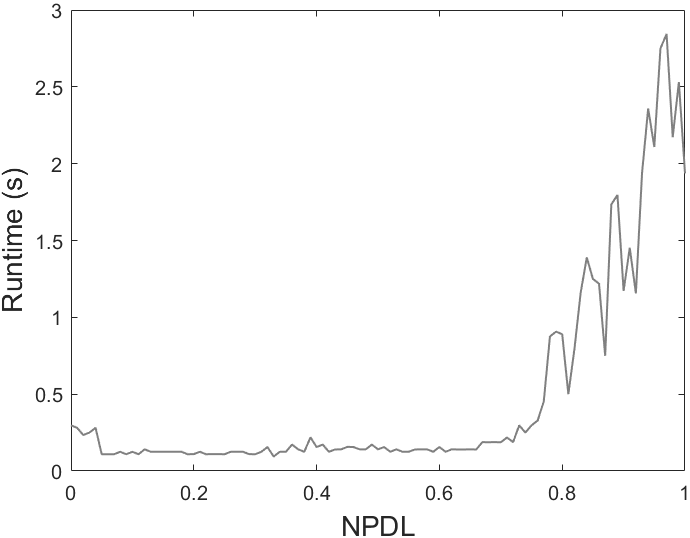}
			\label{fig_ch_dars_t_pdlnpdl}
		}\\
		\subfigure[Simulation 5]{%
			\includegraphics[scale=0.2785]{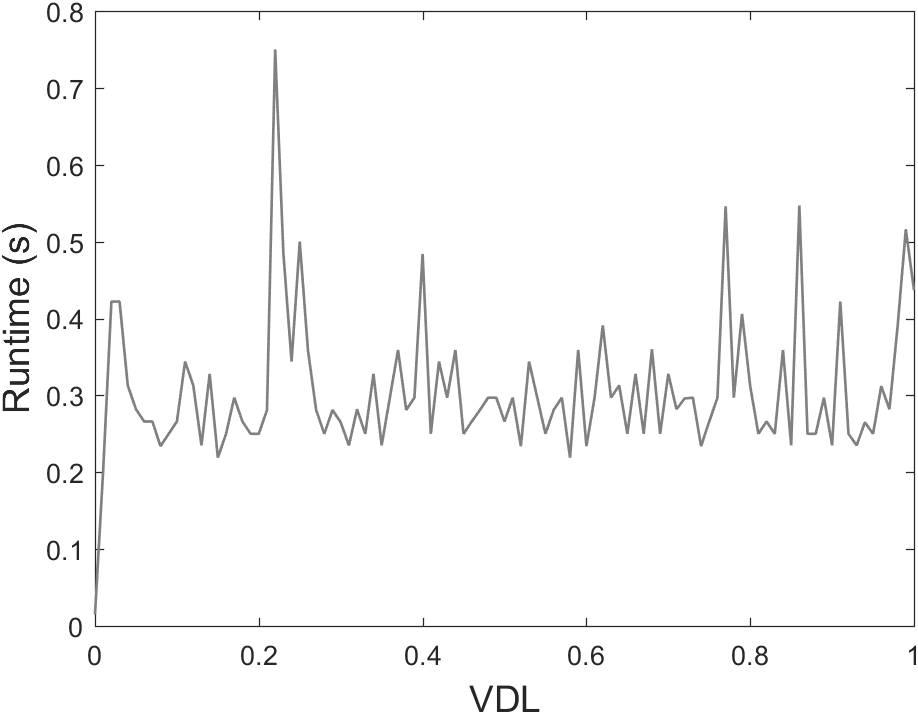}
			\label{fig_ch_dars_t_vdl}
		}
		\subfigure[Simulation 6]{%
			\includegraphics[scale=0.2785]{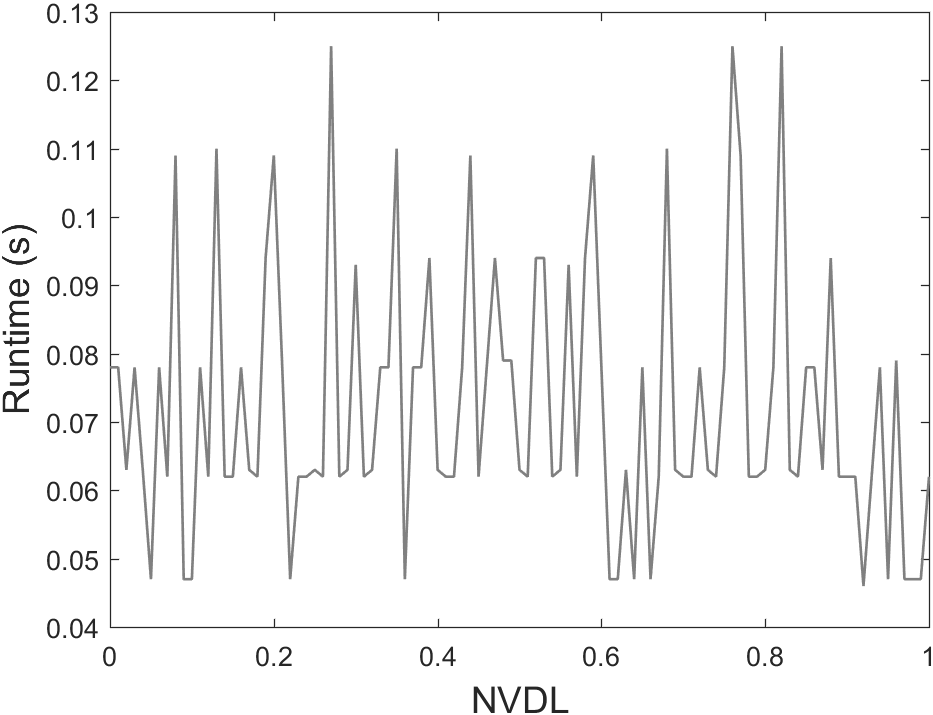}
			\label{fig_ch_dars_t_nvdl}
		}
	\end{center}
	\captionsetup{margin=5ex}
	\caption{%
		Runtime of DARS-ILP.
	}%
	\label{fig_ch_dars_ev}
\end{figure}

(\textbf{RQ7}) is answered by Simulation 1, which evaluates the runtime of DARS-ILP for different numbers of requirements (Figure~\ref{fig_ch_dars_t_n_log}). We observed that increasing the number of requirements increased the runtime of DARS-ILP. Nonetheless, for requirement sets with up to $750$ requirements ($n\leq 750$), the model managed to find the optimal solution in less than a minute. For $750 < n\leq 2000$, the runtime exceeded a minute but did not go beyond two hours. Finally, for $2000 < n \leq  3000$ it took few hours before selection was completed. Simulation 2 demonstrated (Figure~\ref{fig_ch_dars_t_b}) that the runtime of DARS-ILP increased with budget increase (\textbf{RQ8}). The reason is with more budget, more requirements can be selected, which results in a larger solution space; it may take longer to find the optimal subset. 

To answer (\textbf{RQ9}), we simulated requirement selection for various precedence dependency levels (PDLs). Our results (Figure~\ref{fig_ch_dars_t_pdl}) demonstrated that, in general, the runtime of DARS-ILP increased when PDL increased. The reason is increasing PDL limits the number of choices for DARS-ILP as the model needs to respect precedence dependencies; it takes longer for the selection task to complete. Increasing NPDL, on the other hand, had no significant impact on the runtime of DARS-ILP in most places. Nonetheless, for larger NPDLs ($NPDL \rightarrow 1$), runtime was increased. The reason is at such high NPDL, DARS-ILP cannot find a feasible solution with some values due to the high level of conflicts among the requirements. 

Simulation 5 and Simulation 6 were carried out to answer (\textbf{RQ10}) by measuring the runtime of the selection models in the presence of various value dependency levels (VDLs) and negative value dependency levels (NVDLs). Our results demonstrate (Figure~\ref{fig_ch_dars_t_vdl}) that increasing (decreasing) VDL has an inconsistent impact of negligible magnitude on the runtime of DARS-ILP. We further, observed (Figure~\ref{fig_ch_dars_t_nvdl}) that the impact of increasing (decreasing) NVDL on the runtime of DARS-ILP was unpredictable. 

%% file: rq_dars_ilp_scalability.tex
\begin{itemize}[leftmargin=1.8cm]
	\item[(\textbf{RQ7})] How scalable is DARS-ILP to large requirement sets?
	\item[(\textbf{RQ8})] What is the impact of budget on the runtime of DARS-ILP?
	\item[(\textbf{RQ9})] What is the impact of precedence dependencies on the runtime of DARS-ILP?
	\item[(\textbf{RQ10})] What is the impact of value dependencies on the runtime of DARS-ILP?
\end{itemize}

%% file: table_sim_scalability_design.tex
\resizebox {0.36\textwidth }{!}{
\begin{tabular}{lllllll}
	\toprule[1.5pt]
	\rowcolor{gray!30}
	\textbf{\cellcolor{black}\textcolor{white}{Simulation}} &
	\textbf{\cellcolor{black}\textcolor{white}{Size}} &
	\textbf{\cellcolor{black}\textcolor{white}{$\%$Budget}} &
	\textbf{\cellcolor{black}\textcolor{white}{VDL}} &
	\textbf{\cellcolor{black}\textcolor{white}{NVDL}} &
	\textbf{\cellcolor{black}\textcolor{white}{PDL}} &
	\textbf{\cellcolor{black}\textcolor{white}{NPDL}}
	\bigstrut\\
	\hline
	1 &
	[0,3000] &
	50 &
	0.15 &
	0.00 &
	0.02 &
	0.00
	\bigstrut\\ \rowcolor{gray!25}
	2 &
	200 &
	[0,100] &
	0.15 &
	0.00 &
	0.02 &
	0.00
	\bigstrut\\	
	3 &
	200 &
	50 &
	0.15 &
	0 &
    [0,1] &
	0.00
	\bigstrut\\ \rowcolor{gray!25}
    4 &
	200 &
	50 &
	0.15 &
	0.00 &
	0.02 &
	[0,1]
	\bigstrut\\
	5 &
	200 &
	50 &
	[0,1] &
	0.00 &
	0.02 &
	0.00
	\bigstrut\\ \rowcolor{gray!25}
	6 &
	200 &
	50 &
	0.15 &
	[0,1] &
	0.02 &
	0.00
	\bigstrut\\ 
	\bottomrule[1.5pt]
\end{tabular}
}

%% file: conclusion.tex
\section{Conclusions and Future Work}
\label{ch_dars_conclusion}

In this paper, we proposed a fuzzy-based method for integrating value dependencies in software requirement selection. The proposed method, referred to as DARS (Dependency Aware Requirement Selection), uses fuzzy graphs and integer programming to reduce the value loss induced by ignoring value dependencies among software requirements. DARS comprises two major components: (i) a fuzzy-based technique for identifying and modeling value dependencies, and (ii) an integer programming model that explicitly accounts for value dependencies in requirement selection. We have further, proposed a complementary model that can be used to reduce the risk of value loss when quantifying value dependencies is hard. We demonstrated the effectiveness and scalability of DARS by carrying out simulations on the requirements of a real-world software project with results indicative of using DARS reducing value loss. Our results show that (a) compared to existing requirement selection methods, DARS provides higher overall value, (b) maximizing the accumulated value of a requirement subset conflicts with maximizing its overall value -- where value dependencies are considered, and (c) DARS is scalable to large requirement sets with different levels of dependencies among requirements.

The work can be extended by exploring techniques for automated collection of user preferences, which serve as the input to the dependency identification process in DARS. Online application stores can be used for this purpose. Moreover, the definition of value can be extended to human values such as fairness and equality, and variations of DARS can be developed to account for such values as well as the dependencies (conflicts) among them in requirement selection. 